\documentclass[12pt]{iopart}
\usepackage[english]{babel}
\expandafter\let\csname equation*\endcsname\relax
\expandafter\let\csname fl\endcsname\relax
\usepackage{amsfonts,amssymb,amstext,amsthm,amscd,amsbsy}
\expandafter\let\csname endequation*\endcsname\relax
\expandafter\let\csname endfl\endcsname\relax
\usepackage{mathtools}
\usepackage[a4paper,top=1.5cm,bottom=2cm,left=2cm,right=2cm]{geometry}
\usepackage{breakcites}
\usepackage{booktabs}
\usepackage{mathrsfs}
\usepackage{dsfont}
\usepackage{ulem}
\usepackage{graphicx}
\usepackage{subfig}
\usepackage{wrapfig}
\usepackage{indentfirst}
\usepackage{latexsym}
\usepackage{sidecap}
\usepackage{booktabs}
\usepackage{hyperref}

\usepackage{lipsum}
\usepackage{cleveref}
\usepackage[dvipsnames]{xcolor}

\newcommand\myshade{85}
\colorlet{mylinkcolor}{violet}
\colorlet{mycitecolor}{YellowOrange}
\colorlet{myurlcolor}{RoyalBlue}

\hypersetup{
    colorlinks=true,
    linkcolor=myurlcolor,
    citecolor=mycitecolor!\myshade!black,
    filecolor=magenta,      
    urlcolor=magenta,
}

\usepackage{mathrsfs}
\usepackage{textcomp}
\usepackage{braket}
\usepackage{colortbl}
\usepackage{bbm}
\usepackage{comment}
\usepackage{textcomp}
\usepackage{braket}
\usepackage{colortbl}
\usepackage{bbm}
\usepackage{multirow}
\usepackage{vcell}
\usepackage{array}
\usepackage{adjustbox}
\usepackage{tabularx}
\usepackage{enumitem}
\usepackage{musicography}
\usepackage{caption}
\usepackage{tikz}
\usepackage{pgfplots}
\usetikzlibrary{matrix}
\usepackage{tikz}
\usetikzlibrary{shapes.geometric, arrows}

\newcommand{\red}[1]{\textcolor{black}{#1}}

\begin{document}

%\title[The MicrOdosimetry-based modelliNg for RBE ASsessment TOPAS MC extension]{The MicrOdosimetry-based modelliNg for RBE ASsessment (MONAS) TOPAS MC extension: from \textit{in vitro} to clinical RBE calculation}
%\title[]{Integrating in vitro microdosimetric RBE models in clinical practice via the MicrOdosimetry-based modelliNg for RBE ASsessment (MONAS) tool in TOPAS}
\title[Microdosimetric RBE models for particle therapy via the MONAS tool]{Integrating microdosimetric \textit{in vitro} RBE models for particle therapy into TOPAS MC using the MicrOdosimetry-based modeling for RBE Assessment (MONAS) tool}

\author{Giorgio Cartechini$^{1,2}$, Marta Missiaggia$^{1,2}$, Emanuele Scifoni$^{2}$, Chiara La Tessa$^{1,2,3}$ and Francesco G Cordoni$^{2,4*}$}
\address{$^1$Department of Radiation Oncology, University of Miami Miller School of Medicine, 1550 NW 10th Avenue, 33126, Miami (FL), USA}
\address{$^2$Trento Institute for Fundamental Physics and Application (TIFPA), via Sommarive 15, 38123, Trento, Italy}
\address{$^3$Department of Physics, University of Trento, via Sommarive 14, 38123, Trento, Italy}
\address{$^4$Department of Civil, Environmental and Mechanical Engineering, University of Trento, via Mesiano 77, 38123, Trento, Italy}
\ead{francesco.cordoni@unitn.it}
\vspace{10pt}

%\maketitle
\begin{abstract}
\textit{Objective:} In this paper, we present MONAS (MicrOdosimetry-based modelliNg for relative biological effectiveness (RBE) ASsessment) toolkit. MONAS is a TOPAS Monte Carlo extension, that combines simulations of microdosimetric distributions with radiobiological microdosimetry-based models for predicting cell survival curves and dose-dependent RBE.
\textit{Approach:} MONAS expands TOPAS microdosimetric extension, by including novel specific energy scorers to calculate the single- and multi-event specific energy microdosimetric distributions at different micrometer scales. These spectra are used as physical input to three different formulations of the \textit{Microdosimetric Kinetic Model} (MKM), and to the \textit{Generalized Stochastic Microdosimetric Model} (GSM$^2$), to predict dose-dependent cell survival fraction and RBE. MONAS predictions are then validated against experimental microdosimetric spectra and \textit{in vitro} survival fraction data. To show the MONAS features, we present two different applications of the code: i) the depth-RBE curve calculation from a passively scattered proton SOBP by using experimentally validated spectra as physical input, and ii) the calculation of the 3D RBE distribution on a real head and neck patient geometry treated with protons.
\textit{Main results:} MONAS can estimate dose-dependent RBE and cell survival curves from experimentally validated microdosimetric spectra with four clinically relevant radiobiological models. From the radiobiological characterization of a proton SOBP field, we observe the well-known trend of increasing RBE values at the distal edge of the radiation field. The 3D RBE map calculated confirmed the trend observed in the analysis of the SOBP, with the highest RBE values found in the distal edge of the target.
\textit{Significance:} MONAS extension offers a comprehensive microdosimetry-based framework for assessing the biological effects of particle radiation in both research and clinical environments, pushing closer the experimental physics-based description to the biological damage assessment, contributing to bridging the gap between a microdosimetric description of the radiation field and its application in proton therapy treatment with variable RBE.
\end{abstract}

\vspace{2pc}

\submitto{\PMB}
\maketitle

\section{Introduction}
Proton therapy is now widely recognized as an advanced form of radiation therapy compared to the conventional use of photons for treating \red{a steadily increasing number of} types of cancers, especially deep-seated, radioresistant, and hypoxic \cite{loeffler2013charged,tambas2022current}. The advantages of ions over photons are mainly attributed to the localized energy deposition at the end-of-range of the particle, known as the Bragg peak, resulting in a highly conformal dose distribution and normal tissue sparing \cite{durante2017charged}. In addition to the physical advantages, ion \red{beam} therapy is characterized by \red{larger} biological effectiveness. The reason for this is the higher ionization density and more severe damage to \red{cellular} DNA (e.g. double-strand breaks and clustered damage) than photon radiation \cite{scholz2001direct}. The superior biological effectiveness of ions is quantified by the \textit{Relative Biological Effectiveness} (RBE), that is, the ratio between the X-ray (reference) and ion dose causing \red{a} same biological effect \cite{jakel2016icru}. In particle therapy, \red{in general,} to obtain a homogeneous biological effect in the tumor volume, the RBE of ions is included in the treatment plan optimization as a multiplication factor to the absorbed physical dose. In proton treatment planning, an RBE value equal to 1.1 is adopted for both tumor and normal tissue, namely, protons are \red{considered} 10\% more effective than photons. Despite the clinical practice assumes a spatially invariant RBE for protons, pre-clinical \textit{in vitro} and \textit{in vivo} studies have demonstrated that a constant RBE is an oversimplification due to its dependence on numerous parameters (dose, dose rate, cell line, biological endpoint, radiation quality in the specific voxel, etc.) \cite{paganetti2000radiobiological, paganetti2014relative}, with the RBE being significantly above 1.1 in the distal region, \cite{missiaggia2020microdosimetric}. For \red{heavier} ions like Carbon and Helium, the variations in RBE along the beam penetration depth are so significant that a fixed RBE value cannot be deemed appropriate, and current treatment plans are calculated \red{accounting for} a variable RBE, \cite{inaniwa2018adaptation, mairani2022roadmap}.

Therefore, for optimal treatment outcomes that effectively \red{balance the targeting of the tumor and the minimization of the damage} to the surrounding healthy tissue, it is crucial to accurately estimate the RBE \red{at any point of the irradiated field}. The first step in achieving such a challenging goal is to characterize the radiation field, both in terms of macroscopic absorbed energy and the microscopic local pattern of energy deposition. Microdosimetry \cite{zaider1996microdosimetry} has proven over the years to be an extremely powerful tool to accomplish such a task. Microdosimetry is a branch of physics that studies the energy deposition of particles at a scale in the order of a few microns, which is the scale of a cell nucleus, believed to be the most sensitive target to radiation-induced cell killing due to the presence of DNA. At the micron scale, single-particle energy deposition is characterized by large fluctuations due to the inherently stochastic nature of particle interaction, and, therefore, microdosimetry characterizes the radiation field in terms of probability distributions of energy.

By characterizing the energy depositions at the micron scale, microdosimetry provides an ideal tool to link radiation to its biological effects directly. For this reason, many radiobiological models rely on microdosimetry principles, among which the Microdosimetric Kinetic Model (MKM) is the most prominent and widespread in particle therapy \cite{hawkins1994statistical,hawkins1996microdosimetric, inaniwa2018adaptation}. 

%\cite{hawkins1996microdosimetric,zaider1985microdosimetric,kellerer1974theory, inaniwa2018adaptation, scholz2020characterizing, scholz1997computation}. Among these models, the Microdosimetric Kinetic Model (MKM) \cite{hawkins1994statistical,hawkins1996microdosimetric, inaniwa2018adaptation} has emerged as one of the most successful radiobiological models based on microdosimetry.

The MKM is a mechanistic model that predicts the cell survival fraction of irradiated cells based on microdosimetric average values and estimates the resulting RBE \cite{zaider1996microdosimetry}. In particular, the MKM predicts the logarithm of the cell survival fraction of irradiated cells as a linear-quadratic (LQ) function of the imparted dose, \cite{mcmahon2018linear},
\begin{equation}\label{EQN:LQ}
\log S(D) = - \alpha D - \beta D^2\,,
\end{equation}
with $\alpha$ and $\beta$ two radiobiological parameters that depend both on the biological tissue and on \red{the specific} ionizing radiation.
Despite the MKM has displayed notable consistency with experimental data obtained both \textit{in vitro} and \textit{in vivo} \cite{mein2020assessment}, over time, numerous \red{successive} adaptations of the MKM have been developed in the literature, \cite{kase2006microdosimetric,kase2007biophysical,inaniwa2010treatment,sato2006development,bellinzona2021biological}, primarily aimed to address limitations of the original model in specific scenarios where its underlying assumptions were unsuitable. Currently, the MKM represents the standard model used to calculate the RBE in several carbon-ion therapy centers, \cite{mein2020assessment}, and, along with the Local Effect Model (LEM), the MKM is one of \red{the} only two models currently used in clinics for this purpose. Further, the MKM has been used as the reference RBE model in the recently  \red{first} treated patient with helium, \cite{mairani2022roadmap}. The MKM's success and widespread use highlight the importance of microdosimetry in accurately predicting the biological effects of radiation and optimizing treatment planning for patients.

Recently, the \textit{Generalized Stochastic Microdosimetric Model} (GSM$^2$) has been developed, \cite{cordoni2021generalized}, which is a theoretically grounded mechanistic radiobiological model able to include several spatiotemporal stochastic effects inherent to the formation and repair of radiation-induced DNA damage, \cite{cordoni2022cell,cordoni2022multiple,missiaggia2022cell}. GSM$^2$ is a fully probabilistic model that overcomes one of the main assumptions shared by most existing radiobiological models including the MKM, that is the fact that the distribution of the number of damages induced by radiation on DNA is Poissonian. In doing so, GSM$^2$ describes the time evolution of probability distribution of radiation-induced DNA damage rather than focusing on average values as done in the MKM.

Although mechanistic models based on the microdosimetric description of radiation field quality have been shown as an accurate tool for predicting RBE in ion therapy, experimental microdosimetric spectra are still challenging to measure in the daily clinical practice even using commercial detectors, e.g. tissue equivalent proportional counter (TEPC) or silicon-on-insulator (SOI) detector \cite{bradley2001solid,kase2006microdosimetric,rosenfeld2016novel,missiaggia2020microdosimetric,conte2020microdosimetry,missiaggia2021novel,missiaggia2023investigation}. Therefore, numerical algorithms, such as Monte Carlo (MC) particle simulation toolkits, have been demonstrated as a valuable alternative for the microdosimetric characterization of the radiation field \cite{baratto2021microdosimetry, zhu2019microdosimetric, missiaggia2020microdosimetric, missiaggia2023investigation}. Nevertheless, MC simulations for microdosimetric spectra are exceedingly time-consuming, which has prevented their integration into clinical treatment planning systems. This limitation has hindered the utilization of valuable microdosimetric insights within the clinical practice. 
Hence, to expedite computational processes various numerical approximations have been incorporated into MKM models, \cite{inaniwa2018adaptation}, enabling their practical application in everyday scenarios. The most relevant one was the use of an analytical amorphous track model of particle energy deposition at the nanometer and micrometer scale \cite{kiefer1986model, chatterjee1976microdosimetric} which speed up the computation of microdosimetric quantities used as input for the MKM RBE models \cite{kase2007biophysical, inaniwa2018adaptation}.

In this work, we present a novel TOPAS Monte Carlo (MC) microdosimetric extension: MicrOdosimetry-based modelliNg for RBE ASsessment (MONAS). MONAS combines full MC simulations of microdosimetric spectra with clinically relevant microdosimetry-based radiobiological models for cell survival and dose-dependent RBE assessment. 
Furthermore, by utilizing the new scorers of MONAS, it is now possible to generate fully Monte Carlo-based Look-Up Tables (LUTs) of radiobiological parameters $\alpha$ and $\beta$ for RBE-based treatment plan optimization in clinical proton therapy.
MONAS is based on the existing TOPAS microdosimetric extension \cite{zhu2019microdosimetric} which models the main microdosimetry detectors used in literature and scores the lineal energy distributions. The lineal energy $y$ is defined as the energy $\epsilon$ deposited over the target volume mean chord length $l$, i.e. $y = \epsilon/l$. This is the quantity of reference measured in experimental microdosimetry. Starting from the simulated $y$ distributions, we implemented a novel scorer based on specific energy $z$, defined as the energy imparted $\epsilon$ over the mass $m$ of sensitive volume, i.e. $z=\epsilon/m$. The extension allows the user to calculate specific energy distributions at different microscopic scales. These $z$ distributions are the building blocks for the microdosimetry-based radiobiological models implemented in MONAS. We included the GSM$^2$ and three clinically relevant MKM formulations: saturation corrected MKM (MKM-z*) \cite{kase2006microdosimetric}, Double Stochastic MKM (DSMKM) \cite{sato2012cell} and the modified Stochastic MKM (mSMKM) \cite{inaniwa2018adaptation}. Therefore, MONAS predicts dose-dependent cell survival fraction and RBE specifically for the simulated radiation field. 

MONAS simulates experimental microdosimetric distributions acquired with three different detectors, and from validated $z$-distributions, it calculates cell survival curves which can be directly compared with experimental \textit{in vitro} data. Therefore, MONAS is the first full MC toolkit that allows the user to benchmark the physical input and the biological output of the radiobiological models with experiments.
To show that, we compared the MONAS cell survival curves with experimental data from the Particle Irradiation Data Ensemble (PIDE) \cite{friedrich2013systematic}. We also calculated cell survival and RBE depth curves using all radiobiological models available in MONAS for a proton Spread-out Bragg-peak (SOBP), whose microdosimetric spectra were previously measured \red{by us} \cite{missiaggia2023investigation} and used to validate TOPAS.

MONAS is also an accurate and fast tool to predict RBE in proton therapy treatment plans. In particular, in this study, we used the MONAS extension to generate full Monte Carlo-based Look-Up-Tables (LUTs) of radiobiological parameters $\alpha$ and $\beta$ from monoenergetic proton beams. By combining MONAS LUTs and TOPAS MC's precision in tracking therapeutic protons within the patient's anatomy, MONAS allows the calculation of the RBE spatial distribution in a real patient's geometry. In particular, we determined the mixed-filed $\alpha_{mix}$ and $\sqrt{\beta_{mix}}$, defined as the dose-averaged of single particle $\alpha$ and $\sqrt{\beta}$ extrapolated from MONAS LUTs, in each voxel of the scoring mesh. Consequently, we could predict the biological effectiveness of a real proton beam. As an example, we recalculated with TOPAS MC a real head and neck proton treatment, optimized with the Eclipse planning system (Varian Medical Systems, Palo Alto, CA) and delivered at the Dwoskin Proton therapy center (University of Miami). Our study demonstrated that by fully integrating the MONAS toolkit within the TOPAS MC code, we can accurately simulate microdosimetric spectra and use them as input for the microdosimetry base models, enabling us to estimate the effect of radiation on biological tissue in clinical conditions. The present work represents the first crucial step toward bridging the gap between microdosimetry and clinical applications. By providing a comprehensive microdosimetric analysis of the radiation field, we show how real-life clinical and experimental scenarios can greatly benefit, allowing clinicians and researchers to estimate the biological effect of radiation accurately. 

The main contributions of the present paper are:
\begin{description}
\item[(i)] to introduce a Monte Carlo-based microdosimetric toolkit that allows estimating \textit{in vitro} cell survival data and RBE distribution in realistic radiotherapy treatment plans with a full microdosimetric description of radiation.

\item[(ii)] to introduce a new methodology for utilizing the microdosimetry-based radiobiological models to evaluate RBE in real proton therapy treatments.
\end{description}\label{Sec:Introduction}

\section{Material and Methods}

\subsection{Microdosimetric distributions and their moments}\label{SubSec:SpecificEnergy}

Microdosimetry considers two quantities of interest: the \textit{specific energy} $z$ and \textit{lineal energy} $y$ \cite{zaider1996microdosimetry}. 

The \textit{specific energy} $z$ is the ratio between the energy imparted by ionizing radiation $\epsilon$ and the mass $m$ of the sensitive volume,
\[
z = \frac{\epsilon}{m}\,.
\]

The \textit{lineal energy} $y$ is the ratio between by ionizing radiation $\epsilon$ and the mean chord length of the sensitive volume $l$,
\[
y = \frac{\epsilon}{l}\,.
\]

It is possible to relate the lineal to specific energies via the following relation $z=\frac{l}{m}y$. By assuming a spherical site of density $\rho_t=1$ $gcm^{-3}$ and radius $r$ the relations between the $y$ and $z$ is:
\begin{equation}
\label{eq:yToz1}
z = \frac{0.16}{\rho \pi r^2}y
\end{equation}
where $0.16$ is the $Gy-keV$ conversion coefficient.

The lineal energy $y$ is the reference quantity in experimental microdosimetry, whereas the specific energy $z$ is the main quantity of reference in microdosimetry-based radiobiological models. A key difference between lineal energy and specific energy is that the quantity $\epsilon$ refers to the energy imparted in a single event for $y$, or the energy imparted in any number of events for $z$. This implies that when considering the distribution of $z$, the ionization of more than one event must be considered. For this reason, most of the MKM and the GSM$^2$ models are based on the so-called \textit{multi-event} specific energy distributions. 

The \textit{multi-event} distribution is defined starting from the \textit{n-event} distribution $f_n(z)$ which defines the specific energy frequency distribution when exactly \textit{n} events occurred. The n-event distribution can be computed as the n-fold convolution of the single-event specific energy distribution $f_1(z)$:

\begin{equation} 
\begin{split}
f_2(z) & = \int_{0}^{\infty}f_1(z')f_1(z-z')dz' \,,\\
 & ...,  \\
f_n(z) & = \int_{0}^{\infty}f_1(z')f_{n-1}(z-z')dz' \,,\\
\end{split}
\end{equation}

As standard in microdosimetry, the first two moments of the single-event distribution play a crucial role and they are defined as
\begin{align}
\bar{z}_F &= \int_0^{\infty} zf_1(z)dz \,,\label{eq:zF} \\
\bar{z}_D &= \frac{1}{z_F}\int_0^{\infty} z^2f_1(z)dz \,.\label{eq:zD}
\end{align}

The \textit{multi-event} distribution is defined as \cite{bellinzona2021linking,zaider1996microdosimetry},
\begin{equation}
\label{eq:MultiEvent}
    f(z,\lambda_n) = \sum_{n=0}^{\infty}e^{-\lambda_n}\frac{\lambda_n^n}{n!}f_n(z)\,,
\end{equation}
where, by statistical independence of the n events they are assumed to follow the Poisson distribution with mean value $\lambda_n$, so that the term multiplying $f_n$ in equation \eqref{eq:MultiEvent} is the probability of registering exactly n events.

With further computations, \cite{zaider1996microdosimetry}, it is possible to show that the first and second moments of the \textit{multi-event} distribution are:
\begin{align}
    \braket{z} &= \int_0^{\infty} zf(z,\lambda_\nu)dz = \lambda_n z_F \label{eq:MultiEvent1Moment}\,.
%    \braket{z^2} &= \int_0^{\infty} z^2f(z,\lambda_\nu)dz = D^2 + z_DD \label{eq:MultiEvent2Moment} \\
%    \bar{z}_F &= \int_0^{\infty} zf_1(z)dz \label{eq:zF} \\
 %    \bar{z}_D &= \frac{1}{z_F}\int_0^{\infty} z^2f_1(z)dz \label{eq:zD}
\end{align}
Typically, $\braket{z}$ in equation \ref{eq:MultiEvent1Moment} is identified with the absorbed dose D, yielding that the mean value of Poisson distribution of equation \ref{eq:MultiEvent} is given by $\lambda_n=\frac{D}{z_F}$. Further computation, \cite{zaider1996microdosimetry,bellinzona2021linking}, shows
\begin{align}
%    \braket{z} &= \int_0^{\infty} zf(z,\lambda_\nu)dz = \lambda_n z_F \label{eq:MultiEvent1Moment}\\
    \braket{z^2} &= \int_0^{\infty} z^2f(z,\lambda_\nu)dz = D^2 + z_DD \label{eq:MultiEvent2Moment} \,.
%    \bar{z}_F &= \int_0^{\infty} zf_1(z)dz \label{eq:zF} \\
 %    \bar{z}_D &= \frac{1}{z_F}\int_0^{\infty} z^2f_1(z)dz \label{eq:zD}
\end{align}

\subsection{Microdosimetry-based RBE models}\label{SEC:Model}

\subsubsection{The Microdosimetric Kinetic Model and its generalizations}

The MKM is a radiobiological model that utilizes a system of differential equations to predict the survival fraction of irradiated cells. These equations describe the time evolution of the average number of DNA damages, which are divided into two types in the MKM: \textit{sub-lethal} and \textit{lethal lesions}. Sublethal lesions are DNA \red{damages}, typically DNA Double-Strand Breaks (DSB), that can be repaired by the cell, while lethal lesions are unrepairable and result in cell inactivation. To better align with biological data, the MKM postulates that the nucleus is partitioned into sub-units called \textit{domains} so that the number of lethal and sublethal lesions is evaluated for each domain separately. Therefore, the probability of cell survival is estimated by taking into account all the domains into which the nucleus has been divided.

Since its original formulation in \cite{hawkins1994statistical}, the MKM has been widely generalized to include several endpoints and stochastic inter- and intra-cellular effects, \cite{hawkins1996microdosimetric,hawkins2003microdosimetric,inaniwa2010treatment,manganaro2017monte,inaniwa2018adaptation,bellinzona2021linking,attili2022modelling}. We will refrain from providing an exhaustive explanation of all the MKM derivations. Instead, we will focus solely on recalling the formulas for cell survival fraction derived by the models.

Assuming that the reference radiation exhibits an LQ dependence as in equation \eqref{EQN:LQ} between the logarithm of the cell survival fraction and the imparted dose, with given parameters $\alpha_X$ and $\beta_X$, in its original formulation \cite{hawkins1994statistical}, the MKM predicts an LQ cell survival fraction of the form
\begin{equation}\label{EQN:LQ2}
S=e^{-\alpha D -\beta D^2}\,,
\end{equation}
where $\beta=\beta_X$ coincides with the $\beta$ of the reference radiation and $\alpha$ is
\begin{equation}\label{EQN:AMKM}
\alpha = \alpha_0 + \beta \bar{z}_{d,D}\,,
\end{equation}
where $\alpha_0$ is the limit of the $\alpha$ parameter as the LET becomes increasingly small and $\bar{z}_{d,D}$ is the dose-average specific energy defined in equation \eqref{eq:zD} computed over the cell nucleus domain.

A proposed correction to the MKM model, named saturation corrected MKM \cite{kase2006microdosimetric}, aims to improve its alignment with heavy ion data, which \red{exhibits the so-called} \textit{overkill effect}. This effect consists in a decrease of the RBE versus LET, for LET beyond approximately 150 keV/$\mu$m, following its initial raise for increasing LET \cite{kase2006microdosimetric}. In particular, the dose-average specific energy $\bar{z}_{d,D}$ is substituted with its \red{saturation} corrected value, \cite{zaider1996microdosimetry},
\begin{equation}
    \bar{z}_{d,D}^* = \frac{z_0^2\int_0^\infty\!(1-exp[-z_d^2/z^2_0])f_d(z_d)dz_d }{z_{d,F}}\,,
    \label{eq:z*}
\end{equation}
where $z_0$ is the saturation correction parameter for modeling the over-killing effect due to high-LET radiation.

According to the saturation corrected MKM, the $\alpha$ parameters thus becomes
\begin{equation}\label{EQN:AMKM-zStar}
\alpha = \alpha_0 + \beta \bar{z}_{d,D}^*\,,
\end{equation}
whereas the LQ behavior \eqref{EQN:LQ2} remains valid.

Although the saturated corrected MKM shows a better match with experimental data, it still does not include energy deposition variations both at the cell nucleus and at the cell population level. To include these effects, the \textit{DSMKM}, \cite{sato2012cell}, has been proposed. The cell survival fraction at nucleus scale $S_n$ as a function of nucleus-specific energy $z_n$ can be calculated from domain multi-event specific energy distributions as
\begin{equation}
\log S_n(z_n) = -\alpha_0 \int_0^{\infty} \! z'_d f(z_d, z_n/z_{d,F})dz_d -\beta \int_0^\infty \! z'^2_d f(z_d, z_n/z_{d,F})dz_d
\end{equation}
where the domain saturation-corrected specific energy is defined: $z_d'=z_0\sqrt{1-exp(z^2_d/z^2_0)}$. Considering the stochastic nature of $z_n$, the survival fraction of cells irradiated by absorbed dose $D$,
$S(D)$, can be estimated by:
\begin{equation}
    S(D) = \int_0^\infty \! S_n(z_n)f(z_n, D/z_{n,F})dz_n.
    \label{eq:SurvCell}
\end{equation}

The computational requirements for evaluating multi-event functions such as $f(z_d, z_n/z_{d,F})$ and $f(z_n, D/z_{n,F})$ are non-negligible, particularly when combined with macroscopic beam-transport simulations that use Monte Carlo methods to model the irradiation conditions. Therefore, in \cite{sato2012cell} $S_n(z_n)$ formulation has been simplified by omitting the calculation of $f(z_d, z_n/z_{d,F})$, assuming instead that the saturation effect triggered by multiple radiation events to a domain is negligibly small, yielding
\begin{equation}
\log S_n(z_n) = -\alpha_0 z_n\frac{\bar{z}^*_{d,F}}{\bar{z}_{d,F}} -\beta(\bar{z}_{d,D}z_n+z^2_n)\frac{\bar{z}^*_{d,D}}{\bar{z}_{d,D}}
\end{equation}
where the fluence- and dose-averaged specific energy ($\bar{z}_{d,F}$, $\bar{z}_{d,D}$) are calculated according to the equations~\eqref{eq:zF}, ~\eqref{eq:zD}, respectively. The fluence-averaged saturation-corrected specific energy is defined $\bar{z}^*_{d,F}=\int_0^\infty z'_d f(z_d)dz_d$, while the fluence-averaged saturation-corrected specific energy corresponds to equation~\eqref{eq:z*}. This formulation was called stochastic\red{-}MKM (SMKM).

The SMKM has been further simplified in \cite{inaniwa2018adaptation} to speed up the computational time and to be implemented in treatment planning systems. The modified version of SMKM (mSMKM) is based on the assumption that, in charged-particle therapy, the domain-specific energy $z_d$ is in general delivered by a large number of low-energy deposition events, and the events inducing the saturation of complex DNA damages, i.e. events with $z_d > z_0$, are rare. Also, it is assumed that the specific energy imparted $z_n$ is sufficiently close to the macroscopic dose $D$. According to the mSMKM the cell survival fraction is calculated as follows:
\begin{align}
S(D) = exp(-\alpha_{SMK}D-\beta_{SMK}D^2&\left [ 1+D[-\beta_{SMK}+\frac{1}{2}(\alpha_{SMK}+2\beta_{SMK}D)^2\bar{z}_{n,D}] \right ] \\
\alpha_{SMK} &= \alpha_0+\beta \bar{z}^*_{d,D} \,, \qquad \beta_{SMK} = \beta \frac{\bar{z}^*_{d,D}}{\bar{z}_{d,D}}
\end{align}

\subsubsection{The Generalized Microdosimetric Model (GSM$^2$)}

Recently, starting from the building assumptions of the MKM, a novel microdosimetry-based radiobiological model, GSM$^2$, was presented, \cite{cordoni2021generalized, cordoni2022cell,cordoni2022multiple}. GSM$^2$ aims at providing a fully probabilistic model which takes into account the effects of stochasticity in different aspects of radiation-induced damage, e.g., in the initial damage distribution as well as damage evolution. 

Similar to the MKM formulations, GSM$^2$ describes the time evolution of sub-lethal lesion $x$ and lethal lesions $y$ in a cell nucleus, which is divided into smaller sub-domains. Differently from the MKM, GSM$^2$ is able to describe the time evolution of the whole probability distribution of lesions rather than simple average values. 

The survival fraction of a single cell due to the absorption of specific energy $z_n$ in the cell nucleus $S_n(z_n)$ is computed by GSM$^2$ as, \cite{cordoni2022cell},
\begin{equation}\label{EQN:SGSM2}
S_n(z_n) = \Bigl(p^X_0(0|z_n)p^Y_0(0|z_n) + \sum_{x_0=1}^\infty C(x_0)p^X_0(x_0|z_n)p^Y_0(0|z_n) \Bigr)^{N_d} 
\end{equation}
where $p^X_0(x_0|z_n)$, $p^Y_0(0|z_n)$ represent the probability that the domain suffers $x_0$ sub-lesions and 0 lethal lesions, respectively, $N_d$ is the number of domains in which the cell-nucleus is divided and  $C(x_0)$ are weighting terms representing the probability that $x_0$ sub-lethal lesions are
repaired. The terms $p^X_0(x_0|z_n)$, $p^Y_0(0|z_n)$ and $C(x_0)$ can be calculated as follows:
\begin{align}
p^X_0(x|z_n) &= \int_0^\infty \! p^X(x|k z)f(z, z_n/z_{d,F})dz \,,\\
p^Y_0(y|z_n) &= \int_0^\infty \! p^Y(y|\lambda z)f(z, z_n/z_{d,F})dz \,, \\
C(x_0) &= \prod_{x=1}^{x_0} \frac{r x}{(a+r)x + bx(x-1)}\,,
\end{align}
with $p^X(x|k z)$ and $p^Y(y|\lambda z)$ being Poisson distribution of average $k z$ and $\lambda z$, respectively.

Notable enough, it has been shown in \cite{cordoni2022cell,missiaggia2022cell}, that the distributions $p^X_0(x|z_n)$ and $p^X_0(y|z_n)$ can deviate from a Poisson distribution, with particular emphasis on clear differences for sufficiently high LET and doses. Further, $r$, $a$, $b$, $k$, $\lambda$ are model parameters. In particular,  $r$ is the repair rate, $a$ is the rate at which a sub-lethal lesion becomes a lethal lesion and $b$ is the pairwise interaction rate of two sublethal lesions to become a lethal one. At last, $\lambda$ and $\kappa$ are the average yields of lethal and sublethal lesions, respectively, per unit Gray.

Also, another remarkable property of the predicted cell survival fraction \eqref{EQN:SGSM2}, is that it exhibits an LQ behavior at low doses, naturally straightening to a purely linear trend at high doses, \cite{cordoni2022cell,missiaggia2022cell}. This pattern matches empirical evidence collected in cell survival experiments.

At last, as for MKM formulations, the cell survival fraction $S(D)$ of a cell population due to the absorbed dose $D$ can be estimated according to the equation ~\eqref{eq:SurvCell}.

\subsection{TOPAS}

The microdosimetry-based radiobiological extension presented in this work extends the TOPAS MC toolkit \cite{perl2012topas}. TOPAS is an easy-to-use interface to the Geant4 Simulation toolkit \cite{agostinelli2003geant4} allowing both medical physicists and researchers to make Monte Carlo simulations without the necessity of advanced coding knowledge. In \cite{zhu2019microdosimetric} the microdosimetric extension of TOPAS has been implemented, allowing to score of lineal energy $y$ with three different types of detectors: (i) spherical Tissue Equivalent Proportional Counter (TEPC), (ii) a cylindrical TEPC (also known a mini-TEPC) and (iii) Silicon on Insulator (SOI) microdosimeter. These detectors are the reference detectors for microdosimetry, \cite{missiaggia2020microdosimetric,missiaggia2023investigation, de2004mini, bianchi2022topas, bradley2001solid, debrot2018soi}. The microdosimetric extension offers to the user \red{the possibility} to save microdosimetric spectra ($yf(y)$ and $yd(y)$) and the relative average quantities ($y_F$ and $y_D$) for each detector, including the contribution of the particle species of the radiation field. The lineal energy distributions obtained via the TOPAS microdosimetric extension have been benchmarked with experimental data, showing good agreement, \cite{zhu2019microdosimetric,missiaggia2023investigation}.

\subsubsection{MONAS}

\begin{figure}
 \centering
 \includegraphics[width=.8\textwidth]{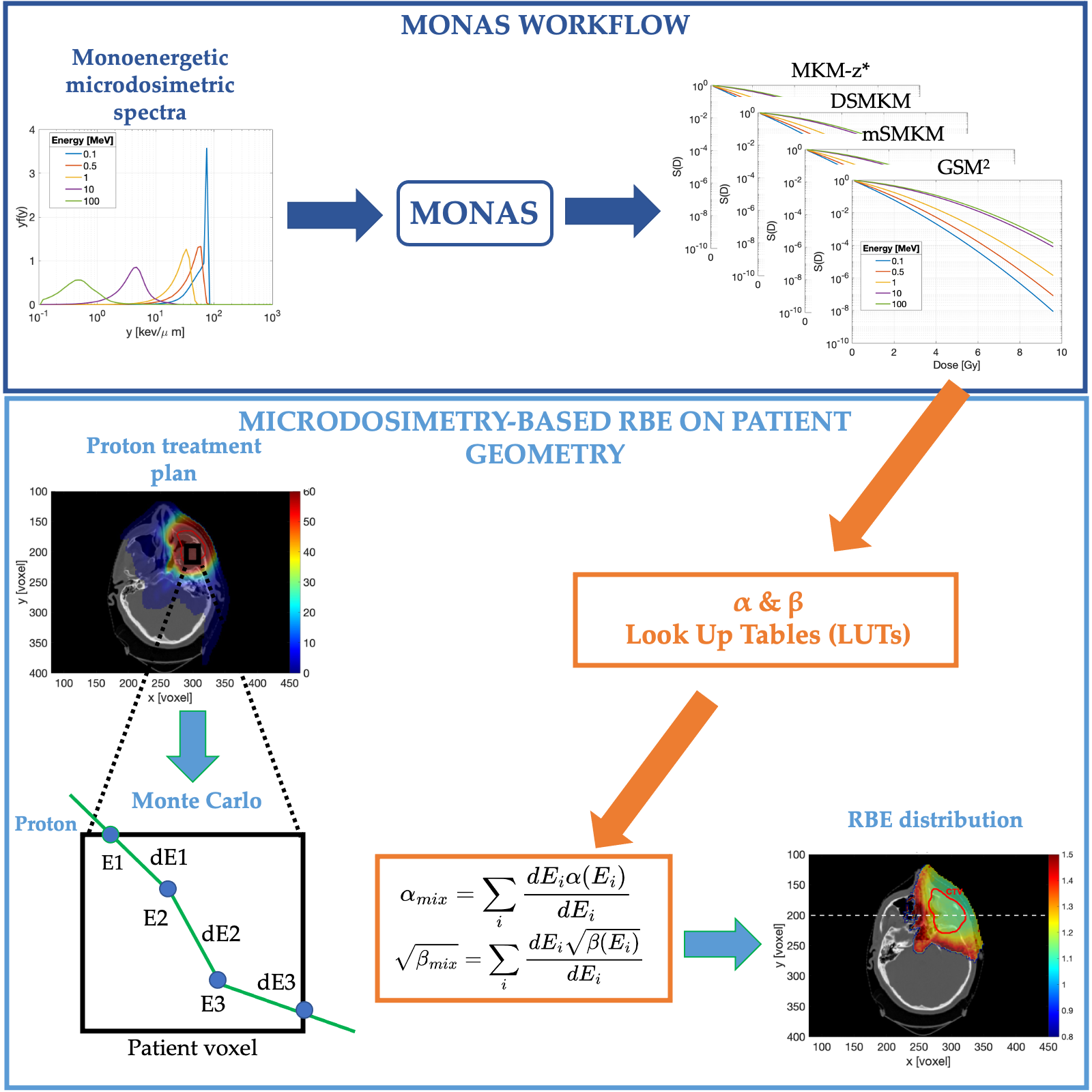}
 \caption{MONAS workflow. Monoenergetic microdosimetric spectra are simulated using TOPAS MC to predict cell survival fraction and RBE with different radiobiological models. The predicted cell survival is thus used to construct $\alpha$ and $\beta$ look-up tables. Then, voxel-based energy depositions are scored and used to reconstruct $\alpha$ and $\beta$ parameters for mixed fields, from which RBE distribution in clinical patients is obtained.}\label{FIG:MONAS}
\end{figure}

MONAS extension starts from the original lineal energy scorer to provide a further toolkit that calculates specific energy $z$, as introduced in Section \ref{SubSec:SpecificEnergy}. Based on specific energy microdosimetric spectra, MONAS predicts cell survival fraction and RBE using the different MKM formulations as well as GSM$^2$ radiobiological model. The workflow of MONAS is depicted in Figure \ref{FIG:MONAS}.

In addition to the parameters of the lineal energy scorer \cite{zhu2019microdosimetric}, new optional parameters were implemented to activate the cell-survival and RBE calculations module according to MKM and GSM$^2$: \textit{GetRBEWithMKModel}, \textit{GetRBEWithGSM2}. The user can further choose one or more MKM formulations available by setting the value of the string parameter \textit{MKMCalculation} equal to i) ``\textit{MKM-z*}" for the saturation corrected MKM, ii) ``\textit{mSMKM}" for the modified-SMKM  or iii) ``\textit{DSMKM}" for the double stochastic MKM. By default, the saturation corrected is set. The cell-survival $S$ is computed as a function of macroscopic absorbed dose $D$, given as input parameter by the user setting the parameter \textit{SurvivalDoses}; thus RBE values are calculated as a function of $S(D)$ as follows
\begin{equation}
RBE(S,D) = \frac{\sqrt{\alpha_X^2-4\beta_X ln(S(D))}-\alpha_X}{2\beta_X D}
\label{eq:RBE}
\end{equation}
where $\alpha_X$ and $\beta_X$ are the Linear-Quadratic coefficients of photon reference radiation. Model-specific radiobiological parameters can be set both for MKM and GSM$^2$ separately, including the reference radiation coefficients for the RBE calculations. Table~\ref{tab:parameters} summarizes all MONAS parameters and their default values. Further details about the evaluation of radiobiological parameters will be given in Section~\ref{sec:RadBioParam}.

The output files include the values set by the user both for the radiobiological model and reference radiation, specific energy spectra both for cell nucleus and domain, cell survival for the macroscopic doses specified, and RBE as a function of cell irradiation dose in ASCII format. The MONAS extension is an open-source code available on GitHub \cite{github}.

% \usepackage{multirow}
\begin{table}[]
\caption{Summary of new input parameters for survival, RBE, and quality factor scorers. The parameter types are indicated according to the TOPAS syntax ($b$ stands for boolean, $sv$ for string vector, $u$ for unitless double, $i$ for integer). Default model-specific biological parameters refer to the HSG cell line.}
\label{tab:parameters}
\begin{adjustbox}{max width=1.1\textwidth,center}
\begin{tabular}{lllp{8cm}}
\toprule
\textbf{Parameter} & \textbf{Type}      & \textbf{Default} & \textbf{Note}                                  \\ \hline
                            &                   &                   &                                               \\
\textit{GetRBEWithMKM}      & b                  & True            & Flag for Survival and RBE calculation with MKM \\
                           &                   &                   &                                               \\
\textit{MKMCalculation} & sv & MKM-z* & String vector with MKM formulations: "MKM-z*", "mSMKM", "DSMKM" \\
\textit{MKM\_Alpha0}        & \multirow{7}{*}{u} & 0.13 Gy$^{-1}$        & \multirow{7}{*}{\parbox{8cm}{MKM parameters}}                   \\
\textit{MKM\_Beta0}          &                    & 0.05 Gy$^{-2}$        &                                                \\
\textit{MKM\_AlphaX}        &                    & 0.19 Gy$^{-1}$        &                                                \\
\textit{MKM\_BetaX}        &                    & 0.05 Gy$^{-1}$        &                                                \\
\textit{MKM\_rho}           &                    & 1 gcm$^{-3}$          &                                                \\
\textit{MKM\_y0}            &                    & 150 keV/$\mu m$       &                                                \\
\textit{MKM\_DomainRadius}  &                    & 0.44 $\mu m$          &                                                \\
\textit{MKM\_NucleusRadius} &                    & 8.0 um             &                                                \\
                           &                   &                   &                                               \\
\textit{GetRBEWithGSM2}      & b                  & false            &  Flag for Survival and RBE calculation with GSM$^2$\\
\textit{GSM2\_AlphaX}       & \multirow{10}{*}{u} & 0.19 Gy$^{-1}$        &   \multirow{10}{*}{\parbox{8cm}{GSM$^2$ model parameters}}    \\
\textit{GSM2\_BetaX}         &                    & 0.05 Gy$^{-2}$        &                                                \\
\textit{GSM2\_rho}          &                    & 1 g/cm$^3$          &                                                \\
\textit{GSM2\_a}       &                    & 0.037 h$^{-1}$           &                                               \\
\textit{GSM2\_b}       &                    & 0.182 h$^{-1}$           &                                               \\
\textit{GSM2\_r}       &                    & 3.641 h$^{-1}$           &                                           \\
\textit{GSM2\_DomainRadius}       &                   & 0.8 $\mu m$           &                                           \\
\textit{GSM2\_NucleusRadius}       &                    & 5.0 $\mu m$             &                                           \\
                           &                   &                   &                                               \\
\textit{SetMultieventStatistic}       & i                   & $10^4$          &  Number of iterations for calculating the \textit{multi-event} probability distribution via Monte Carlo approach \\
                           &                   &                   &                                               \\
\textit{SaveSpecificEnergySpectra}   & b                  & false            &  Flag for saving in an ASCII file the \textit{single-} and \textit{multi-event} distributions for cell domain and nucleus \\
                           &                   &                   &                                               \\
%\textit{QualityFactorCalculation}& sv                  & ICRU40            & String vector with quality factor formulations: "ICRU40", "KellererHahn" \\ 
\bottomrule
\end{tabular}
\end{adjustbox}
\end{table}

\subsubsection{Specific Energy Spectra}\label{SEC:MMEn}
Parallel to the default lineal energy scorer, MONAS includes the calculation of specific energy quantities converting \textit{single-event} lineal energy $y$ into \textit{single-event} specific energy $z_1$ according to the equation 
 \eqref{eq:yToz1}.
\textit{Single-event} and \textit{multi-event} specific energy spectra are calculated and used for cell survival and RBE evaluation, as described in Section \ref{SEC:Model}. By setting the boolean parameter \textit{SaveSpecificEnergySpectra}, the user can save in ASCII format  the \textit{single-event} and \textit{multi-event} distributions calculated on cell domain and cell nucleus: $z_{x}$, $f_{x,1}(z_{x})$, $z_{x}f_{x,1}(z_{x})$, $z_{x}f_{x}(z_{x}, D/z_{xF})$, where the subscript $x$ can be either $d$ for the domain and $n$ for the nucleus distributions.

\begin{figure}
    \centering
    \includegraphics[width=\textwidth]{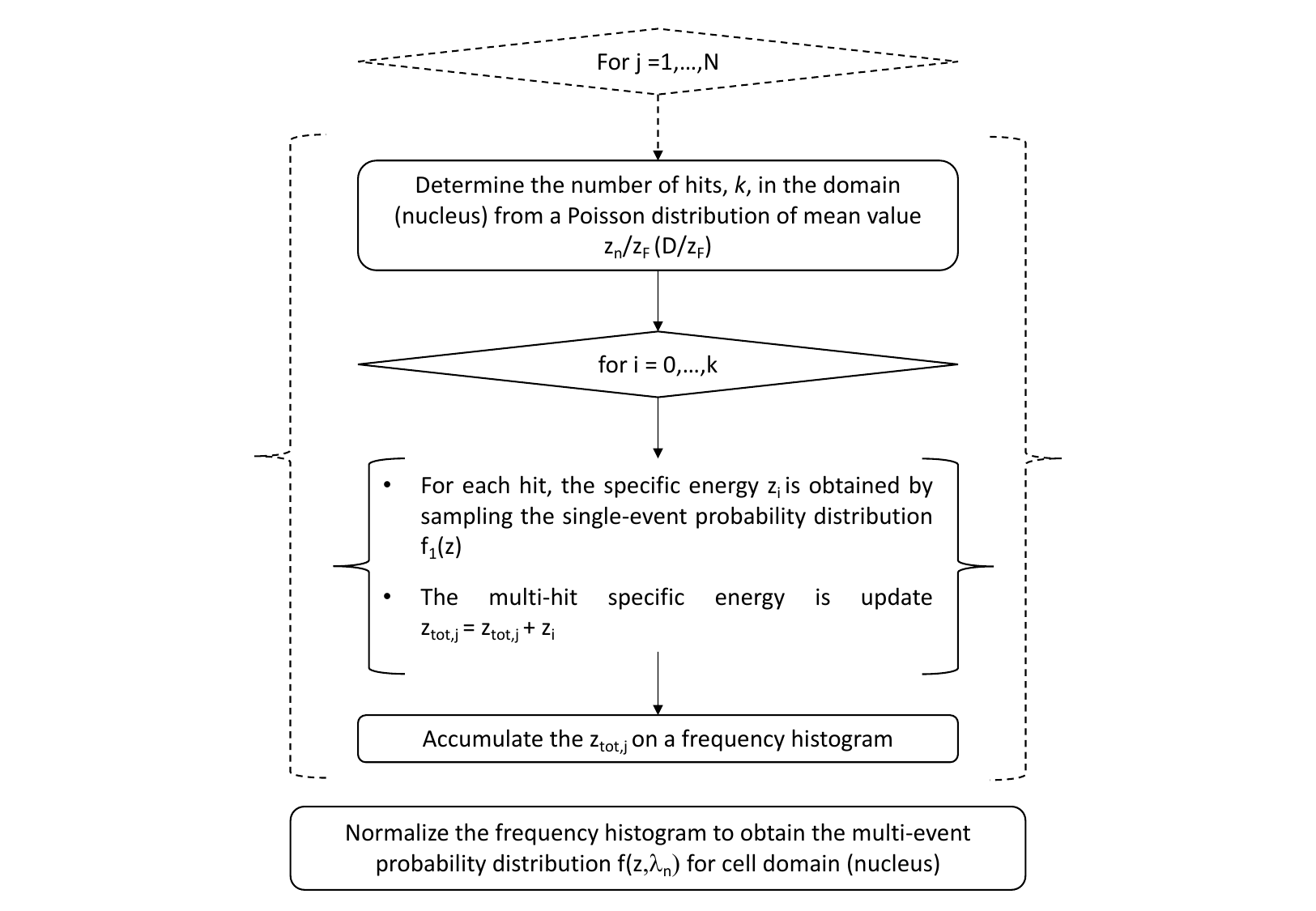}
    \caption{Flowchart of \textit{multi-event} probability distribution calculation with a Monte Carlo approach by sampling distributions via random number generators. Steps inside curly brackets are the body of the two for loops \red{iterating} on the number of \textit{multi-event} statistics and cell domain (nucleus) particle hits.}
    \label{fig:flowchart}
\end{figure}

Since the n-fold convolution for the \textit{multi-event} calculation is time-consuming, especially for high doses when the number of convolutions increases, a Monte Carlo approach for evaluating the \textit{multi-event} distribution has been specifically implemented according to the following workflow: (i) the number of tracks $k$ which deposit energy on sensitive volume is generated from a Poisson distribution with mean value $\lambda_n=z_n/z_{d,F}$ for cell domain and $\lambda_n=D/z_{n,F}$ for nucleus, respectively. Then, (ii) $k$ \textit{single-event} specific energies $z_1$ are sampled from the single-event probability distribution $f_1(z)$ and summed up to obtain the total specific energy $z_{tot}=\sum_{i=1}^k z_{1,i}$ deposited in the target; (iii) the \textit{multi-event} probability distribution $f(z,\lambda_n)$ both for cell domain and the nucleus is thus constructed by iterating steps (i) and (ii) N-times according to the parameter \textit{SetMultieventStatistic}. A scheme of the algorithmic construction described is depicted in figure \ref{fig:flowchart}.

\subsection{Cell-survival and RBE}

The main feature of the MONAS extension is the prediction of cell survival curves and dose-dependent RBE after irradiation with ion beams. To prove the accuracy of the toolkit, we determined the radiobiological parameters specific to each model by comparing the survival predictions with experimental \textit{in vitro} data on the Human Salivary Gland (HSG) cell line. Once the model parameters were determined, we used MONAS to predict RBE distribution in two relevant cases for proton therapy: passively scattering proton Spread-Out Bragg Peak (SOBP) \cite{missiaggia2023investigation,tommasino2019new} and head and neck proton therapy irradiation delivered at the Dwoskin Proton therapy center (University of Miami) and optimized with the Eclipse treatment planning system (Varian Medical Systems, Palo Alto, CA). 

\subsubsection{Radiobiological parameters} \label{sec:RadBioParam}

The radiobiological models implemented in this work are based on a variable number of free parameters which are independent of the radiation type, but they are only cell-line dependent. 

These parameters were estimated by fitting the models with \textit{in vitro} cell-survival experimental curves available in the literature and measured for a specific irradiation condition. It is worth stressing that, previous works \cite{kase2006microdosimetric,sato2012cell,inaniwa2018adaptation} reported MKM parameters for HSG cell line. Nonetheless, since the physical estimation of the radiation field is different from what was implemented in MONAS, the biological parameters have been fitted to ensure the highest reproducibility of \textit{in vitro} cell survival experiments. In the present work, we characterized the radiation field in terms of lineal energy and specific energy spectra by exploiting the TOPAS toolkit, a condensed history Monte Carlo algorithm. On the contrary, a different methodology was used both for the DSMKM and for the modified-SMKM \cite{sato2012cell, inaniwa2013effects}. In fact, DSMKM exploits a combination of microdosimetric Monte Carlo simulations with PHITS code \cite{sato2013particle} in 1 $\mu m$ volume for the cell domain \cite{sato2009biological,sato2006development}, while a Fermi function of macroscopic LET is used for evaluating the specific energy probability distribution on cell nucleus \cite{sato2012cell}. The modified-SMKM, instead, exploits analytical amorphous-track models to describe the radial dose distribution of the ion track and, thus, to compute the specific energy deposited on the cell domain and nucleus ~\cite{chatterjee1976microdosimetric, kiefer1986model, kase2006microdosimetric, inaniwa2010treatment}.
Due to the intrinsic difference in the description of radiation energy deposition in the sensitive volume between TOPAS and the previous work, we re-calculated the model parameters. A systematic comparison between Monte Carlo condensed history (e.g. TOPAS MC) and track structure algorithm for the calculation of specific energy spectra is out of the scope of this work.

Therefore, the parameters for the MKM and GSM$^2$ were determined to reproduce \textit{in vitro} experimental data of HSG cells. Experimental cell survival curves were taken from the Particle Irradiation Data Ensemble (PIDE) \cite{friedrich2013systematic} in which a large amount of cell survival data are systematically collected and analyzed as a function of particle LET, cell line, and reference radiation. Regarding the estimation of specific energy spectra, we simulated with TOPAS MC (v3.7) a water sphere placed in a vacuum world to avoid particle energy loss outside the sensitive volume. The sphere was irradiated by a mono-energetic $^{3}He$ (10.2 MeV/u and 4.89 MeV/u) and $^{12}C$ (126 MeV/u and 126 MeV/u) ion beams as reported in the experiments \cite{furusawa2000inactivation, friedrich2013systematic}. The beam was modeled as the \textit{Environment} particle source available in TOPAS. It creates an isotropic, uniform radiation field enclosing the water sphere. The default TOPAS physics list was used \cite{jarlskog2008physics}. Then, we scored the specific energy spectra in the sensitive volume.

\subsubsection{Cell survival and RBE for a proton Spread-out Bragg-peak}\label{SEC:MMSOBP}

Recently in \cite{missiaggia2023investigation} a systematic characterization of the radiation field produced by passively scattered proton SOBP generated by a pencil beam of 148 MeV has been presented. Microdosimetric spectra were acquired with spherical TEPC both in-field and out-of-field. The same work also shows a good agreement between TOPAS microdosimetric simulations and experiments.

Starting from validated lineal energy spectra, we calculated cell survival, and RBE along the beam axis using the MONAS code. As described in \cite{missiaggia2023investigation}, we simulated the Far West Technologies LET-1/2 spherical TEPC with an active volume made of pure propane gas ($C_3H_8$) at an operative pressure such that it is equivalent to a tissue sphere of 2 $\mu m$ diameter. The default physics list was used and the particle production cut was set according to the microdosimetric extension \cite{zhu2019microdosimetric}. About $10^7$ primaries were simulated for each position along the beam axis. 200 kV X-ray was chosen as reference radiation for the HSG cell line, with $\alpha_X=0.19$ $Gy^{-1}$ and $\beta_X=0.05$ $Gy^{-2}$ \cite{kase2006microdosimetric}.

\subsubsection{Cell survival and RBE assessment in a patient case}
Predicting the cell survival fraction and RBE within the patient's anatomy involved a two-step process. First, we utilized the MONAS extension to create lookup tables (LUTs) of radiobiological parameters $\alpha$ and $\beta$ for each model and proton beam energy. Then, we incorporated the MONAS LUTs into the TOPAS Monte Carlo particle transport algorithm to score the survival fraction and RBE in each voxel of the patient.
To construct the LUTs, we irradiated a sphere of $1$ $\mu m$ of radius made of water with monoenergetic proton beams at different energies (from 0.1 MeV to 300 MeV) and we scored the cell survival fraction with the MONAS extension. We fitted each survival curve with the LQ model and we determined the $\alpha$ and $\beta$ coefficients. Using the TOPAS particle transport algorithm, we created a novel scorer to calculate the mixed filed $\alpha_{mix}$ and $\beta_{mix}$ for each voxel of the patient scoring mesh starting from the LUTs of monoenergetic beams. In particular, for each step $s$ of primary and secondary protons inside the voxel $v$, we registered the kinetic energy of the particle ($E_{v,s}$) and the energy deposited along the step plus the energy released to secondary $\delta$-rays ($dE_{v,s}$) \cite{cortes2015critical}. As standard, the $\alpha_{mix}$ and $\beta_{mix}$ are thus calculated as a weighted sum of $\alpha$ and $\beta$ for the monoenergetic beam, \cite{zaider1980synergistic},
\begin{align}
    \alpha_{mix} &= \frac{\sum_{s=1}^{N_s}dE_{v,s}\alpha(E_{v,s})}{\sum_{s=1}^{N_s}dE_{v,s}}\\
    \sqrt{\beta}_{mix} &= \frac{\sum_{s=1}^{N_s}dE_{v,s}\sqrt{\beta(E_{v,s})}}{\sum_{s=1}^{N_s}dE_{v,s}}\\
\end{align}

To show the applicability of this approach in real patient irradiation, we simulated in TOPAS a head and neck treatment plan, optimized with the Eclipse treatment planning system (Varian Medical Systems, Palo Alto, CA) at the Dwoskin Proton therapy center (University of Miami). The plan is composed by 2 coplanar fields at 30 and 60-degree gantry angles and a third noncoplanar at 30 degrees gantry and couch angle at 300 degrees. All fields employed a range shifter of 57 mm of water equivalent thickness. A uniform biological dose of 60 Gy(RBE) in 30 fractions was prescribed to the target using a constant RBE equal to 1.1.

We estimated the cell survival fraction and the dose-dependent RBE with the approach described above. The radiobiological parameters for the reference radiation were $\alpha_X = 0.19$ $Gy^{-1}$ and $\beta_X = 0.05$ $Gy^{-2}$ for HSG cell line. 

To compare our findings, we repeated the same analysis using the mSMKM based on the amorphous track description of radiation energy description at the microscopic scale \cite{inaniwa2018adaptation}, which is the version currently used in carbon ion therapy. To generate the dose-averaged specific energies per event $\bar{z}_{d,D}$ and $\bar{z}^{*}_{d,D}$ imparted to the domain, and to the cell nucleus, $\bar{z}_{n,D}$, we employed the \textit{Survival toolkit} code \cite{manganaro2018survival}. The resulting cell survival fraction and RBE are as described in \cite{inaniwa2018adaptation}.

\label{Sec:MM}
\section{Results}\label{SEC:Res}

\subsection{Radiobiological parameters}

\begin{figure}
 \centering
\subfloat[]{
	\begin{minipage}[c][\width]{
	   0.49\textwidth}
	   \centering
	   \includegraphics[width=1\textwidth]{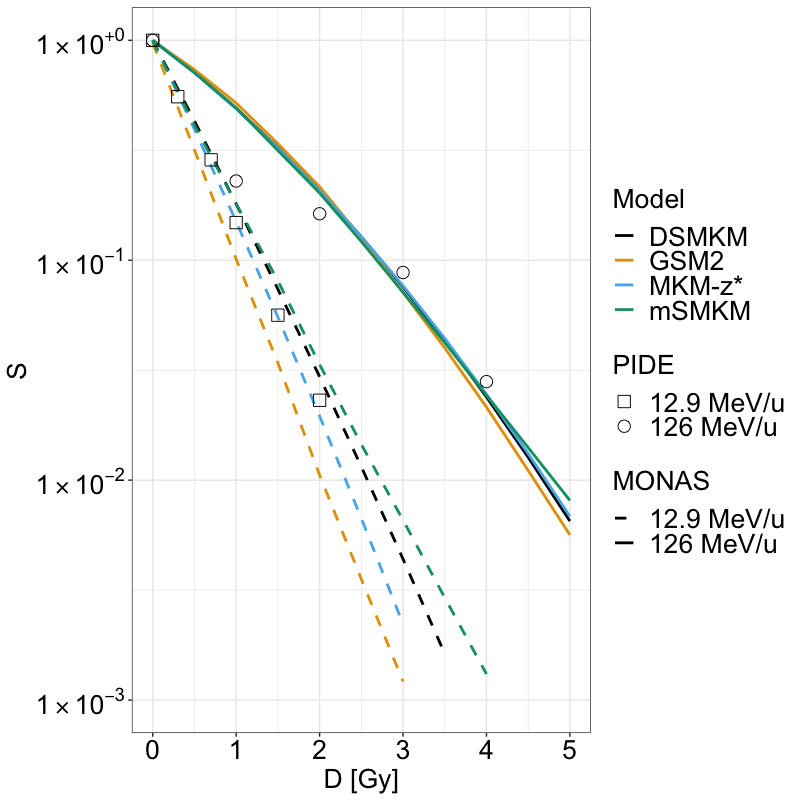}
	\end{minipage}}
	\hfill 
	\subfloat[]{
	\begin{minipage}[c][\width]{
	   0.49\textwidth}
	   \centering
	   \includegraphics[width=1\textwidth]{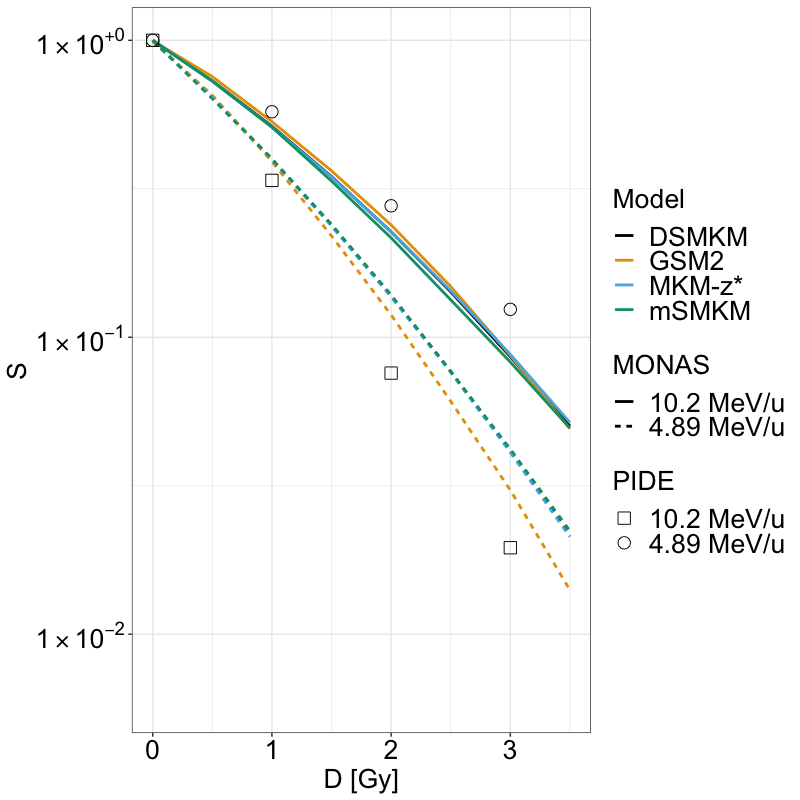}
	\end{minipage}}
	\hfill 
	\caption{Survival fraction of HSG cellls irradiated with $^{12}C$ at 12.9 MeV/u and 126 MeV/u (panel (a)), and $^{3}He$ at 10.2 MeV/u and 4.89 MeV/u (panel (b)). Experimental data from \cite{furusawa2000inactivation}, tabulated in the PIDE.}
	\label{fig:HSGSurvival}
\end{figure}

Table \ref{tab:HSGCell} reports the MKM and GSM$^2$ parameters that give the best fit for the \textit{in vitro} cell-survival data of the HSG cell line when irradiated with $^{12}C$ and $^{3}He$ ion beams. Figure \ref{fig:HSGSurvival} shows the experimental cell survival fraction, as taken from the PIDE dataset, \cite{friedrich2013systematic}, for the HSG cell line, compared to the corresponding predicted cell survival curves using the four radiobiological models with parameters as given in Table \ref{tab:HSGCell}. All radiobiological models are in good agreement with the HSG \textit{in vitro} cell survival curves for both carbon and helium ions.\\ 

Regarding the fitted MKM parameters, given a new description of the physics of the radiation field employed in this study, all parameters have been recalibrated due to a discrepancy between cell survival experimental results and prediction using parameter values reported in the original papers. In particular, in \cite{kase2006microdosimetric}, the MKM-z* parameters extrapolated directly from microdosimetric measurements of $y_D$ and from \textit{in-vitro} HSG cell survival data are reported to be $\alpha_0=0.13$ $Gy^{-1}$, $\beta=0.05$ $Gy^{-2}$. Absolute values of $\alpha_0$ and $\beta$ do not agree with our fit, nevertheless, the $\alpha/\beta$ ratio and domain radius are consistent with \cite{kase2006microdosimetric}. Further, DSMKM and mSMKM parameters are different from the originals, \cite{sato2012cell,inaniwa2018adaptation}, but nevertheless, the difference in absolute value is moderate. 

At last, concerning GSM$^2$, coherently with \cite{missiaggia2022cell}, only $a$, $b$, and $r$ parameters were fitted whereas domain and cell nucleus radius were set a priori. This is done to avoid overfitting.

\begin{table}
\centering
\begin{tabular}{lccc|cc}
\multicolumn{6}{c}{\textbf{HSG}} \\
\toprule
& \multicolumn{1}{c}{\textbf{MKM-z*}} & \multicolumn{1}{c}{\textbf{DSMKM}} & \multicolumn{1}{c}{\textbf{mSMKM}} & &\multicolumn{1}{c}{\textbf{GSM$^2$}} \\ \midrule
$\alpha_0$ $[Gy^{-1}]$      & 0.19                                         & 0.16 & 0.16                               & $a$ $[h^{-1}]$                 & 0.037           \\
$\beta_0$ $[Gy^{-2}]$        & 0.07                     & 0.08  & 0.08                        & $b$ $[h^{-1}]$                 & 0.182          \\
$y_0$ $[keV/\mu m]$             & 150                                      & -  & -                                & $r$ $[h^{-1}]$                 & 3.641            \\
$R_d$ $[\mu m]$         & 0.44            & 0.46 & 0.46                                                       & $R_d$ $[\mu m]$                 &  0.80           \\
$R_n$ $[\mu m]$        & -                                        & 8.0 & 8.0                                & $R_n$ $[\mu m]$           &  5.0            \\
\bottomrule
\end{tabular}
\caption{MKM and GSM$^2$ model parameters, as reported in Section \ref{SEC:Model}, for Human Salivary Gland cell line. Experimental data are taken from the PIDE dataset, \cite{friedrich2013systematic}.} 
\label{tab:HSGCell}
\end{table}

\subsection{Specific energy spectra}
\begin{figure}
 \centering
\subfloat[]{
	\begin{minipage}[c][.7\width]{
	   0.49\textwidth}
	   \centering
	   \includegraphics[width=1\textwidth]{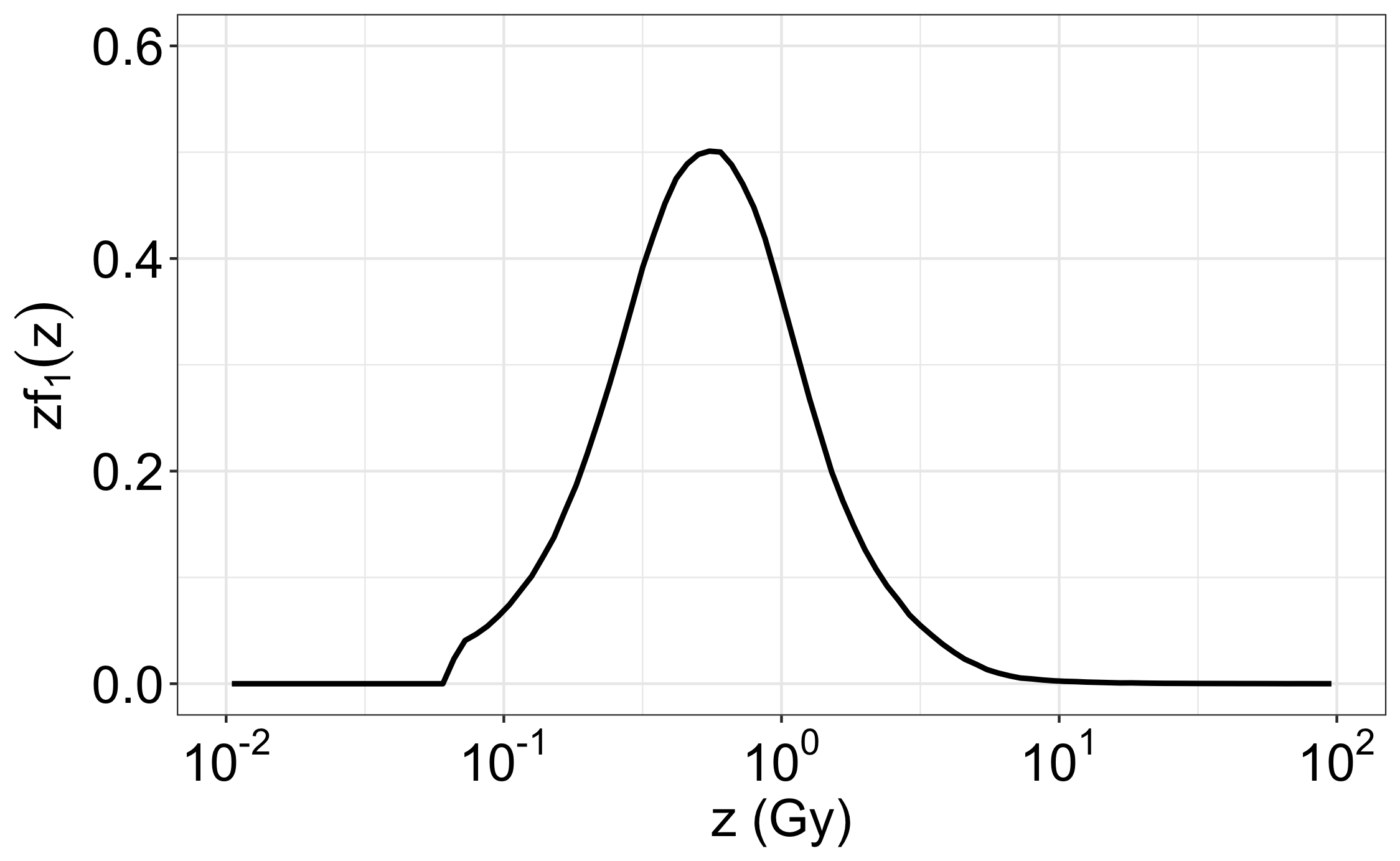}
	\end{minipage}}
	\hfill 
	\subfloat[]{
	\begin{minipage}[c][.7\width]{
	   0.49\textwidth}
	   \centering
	   \includegraphics[width=1\textwidth]{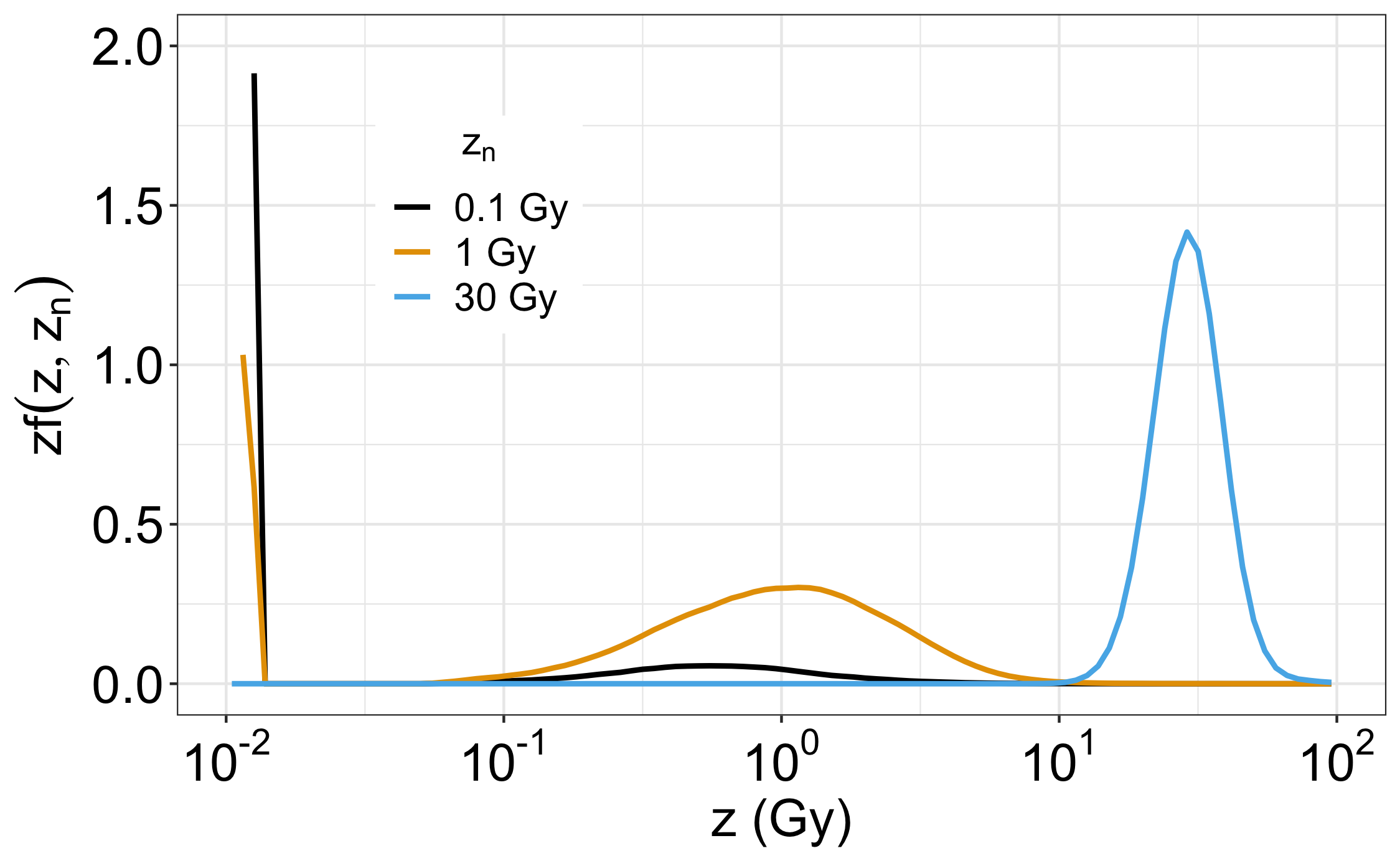}
	\end{minipage}}
	\hfill 
	\caption{\textit{Single-event} (a) and \textit{multi-event} (b) specific energy probability distributions calculated on cell domain of radius 0.46 $\mu m$. The spectra were scored by locating the sensitive volume at the center of a proton SOBP.}
	\label{fig:SpecifiEnergySpectraDomain}
\end{figure}

The novel specific energy scorer implemented in this work allows calculating the \textit{single-} and \textit{multi-event} probability distributions in equation \eqref{eq:MultiEvent}, both at the domain and cell nucleus scales, that is, in the order of 1 micron and 10 microns, respectively.
Figure \ref{fig:SpecifiEnergySpectraDomain} shows the \textit{single-event} $zf_1(z)$ and \textit{multi-event} $zf(z,z_n)$ distributions computed at different averages doses $z_n$=0.1, 1 and 30 Gy for a cell domain radius equal to 0.46 $\mu m$, chosen as the representative domain sizes for the DSMKM and the mSMKM. These spectra were calculated by locating the TEPC active volume at the center of the SOBP proton beam described in Section \ref{SEC:MMSOBP}.

The single-event distribution $zf_1(z)$ preserves a similar shape as the corresponding $yf(y)$ distribution, as could be expected from equation \eqref{eq:yToz1}. On the contrary, the shape of the \textit{multi-event} distribution changes significantly as a function of the dose delivered to the cell domain $z_n$. In particular, to lower doses, e.g. $z_n$=0.1 and 1 Gy, it corresponds a higher probability of null energy deposition in the domain due to \red{the case of} no tracks hitting the target. This phenomenon is evident by the high peak at $z=0$ Gy. As the dose $z_n$ increases, since the average value of Poisson distribution is proportional to $z_n$, the probability to score zero tracks vanishes, and consequently the peak at null $z$ disappears. All three multi-event distributions have an average value equal to $z_n$, and furthermore, at higher $z_n$ the distribution is uni-modal and peaked at $z=z_n$ with a Gaussian-like shape.

\begin{figure}
 \centering
    \includegraphics[width=.8\textwidth]{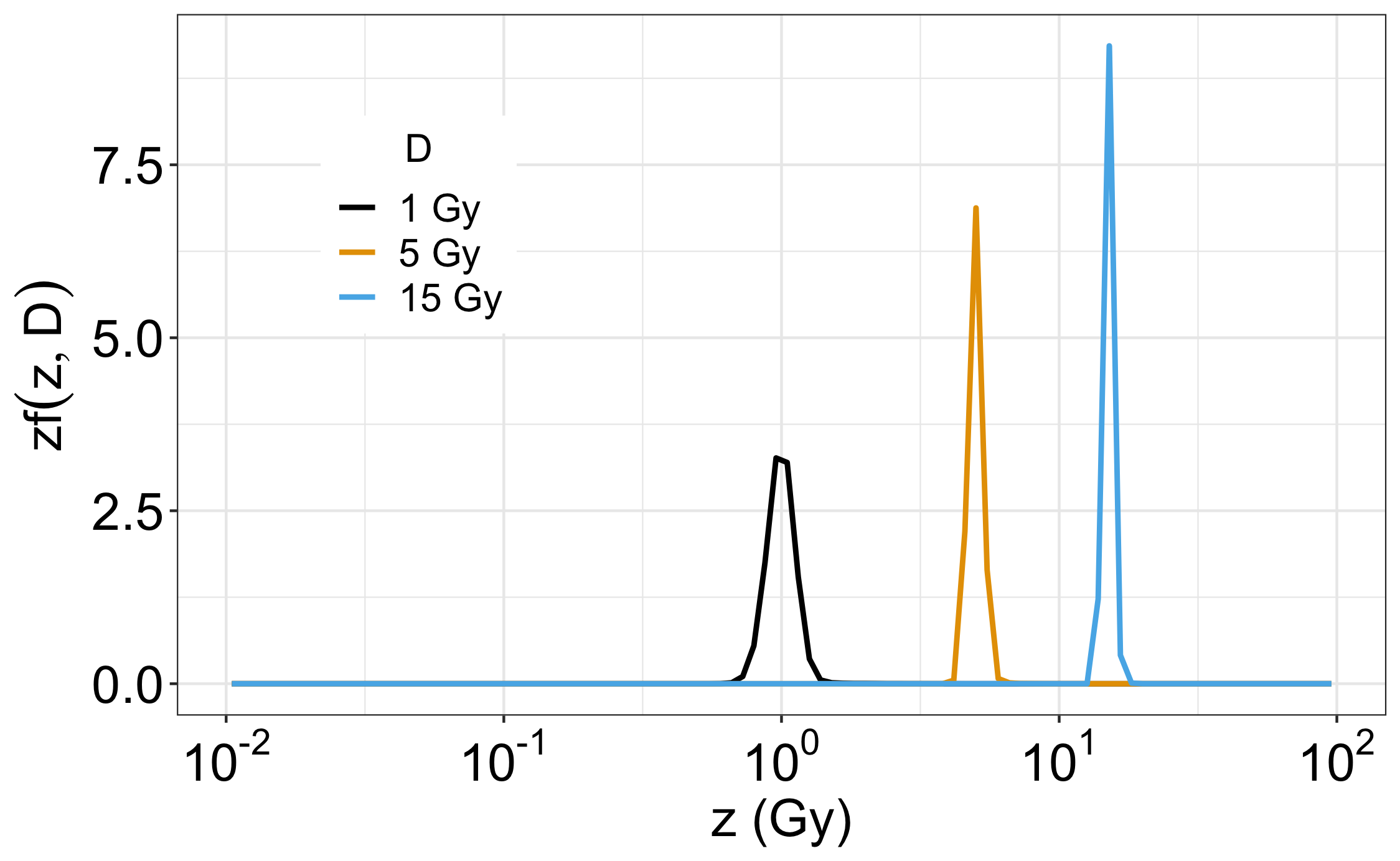}
	\caption{\textit{Multi-event} specific energy probability distributions calculated on cell nucleus of radius 8 $\mu m$ at $D=$ 1, 5, 15 Gy.}
	\label{fig:SpecifiEnergySpectraCell}
\end{figure}

The $zf(z,D)$ distribution at the cell nucleus scale, with a radius of 8 $\mu$m, as a function of macroscopic dose $D=$1, 5, 15 Gy is shown in Figure \ref{fig:SpecifiEnergySpectraCell}. As before, the spectrum is generated by an SOBP proton beam. The \textit{multi-event} distribution has a uni-modal Gaussian-like shape peaked at $z=D$ even at a low dose of $D=1$ Gy. 

%%%%%%%%%%%%%%%%%%%%%
%       SOBP        %
%%%%%%%%%%%%%%%%%%%%%
\subsection{Cell survival and RBE for a proton Spread-out Bragg-peak}\label{sec:RES_SOBP}

Cell survival fraction and RBE were calculated using MKM formulations and GSM$^2$ along the beam axis using experimentally validated microdosimetric spectra, \cite{missiaggia2023investigation}, as described in Section \ref{SEC:MMSOBP}. The physical 3D dose distribution was simulated in the water phantom using a voxel size of $1 \times 1 \times 1$ $mm^3$ and normalized to 1.8 Gy at the center of SOBP to obtain a biological dose of 2 Gy(RBE) for constant RBE equal to 1.1.

\begin{figure}
 \centering
 \subfloat[]{
	\begin{minipage}[c][.8\width]{
	   0.49\textwidth}
	   \centering
	   \includegraphics[width=1\textwidth]{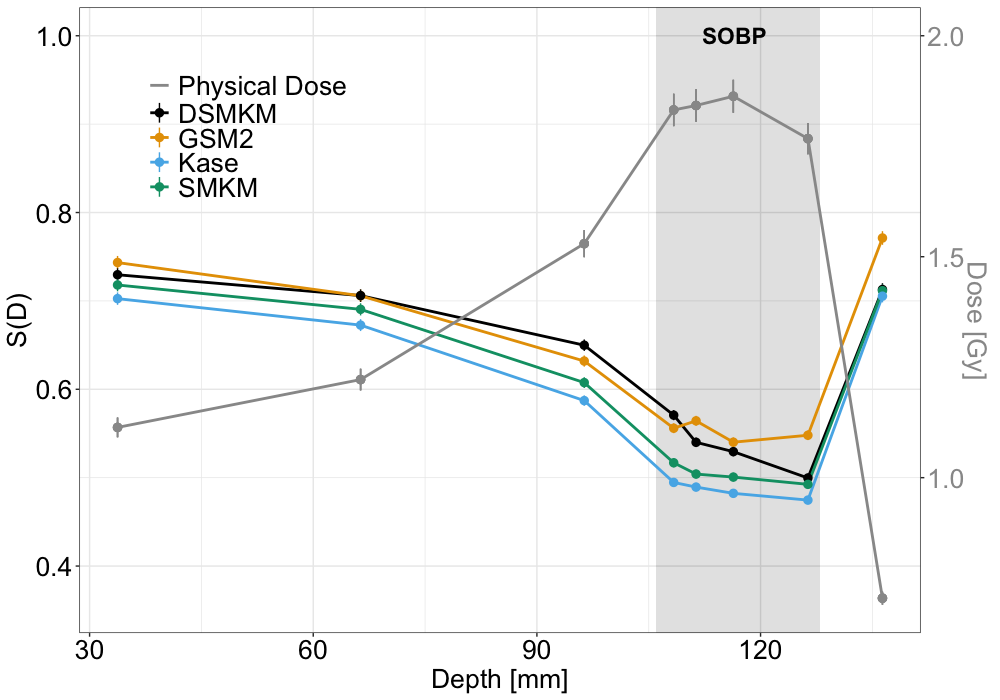}
	\end{minipage}}
	\hfill 
	\subfloat[]{
	\begin{minipage}[c][.8\width]{
	   0.49\textwidth}
	   \centering
	   \includegraphics[width=1\textwidth]{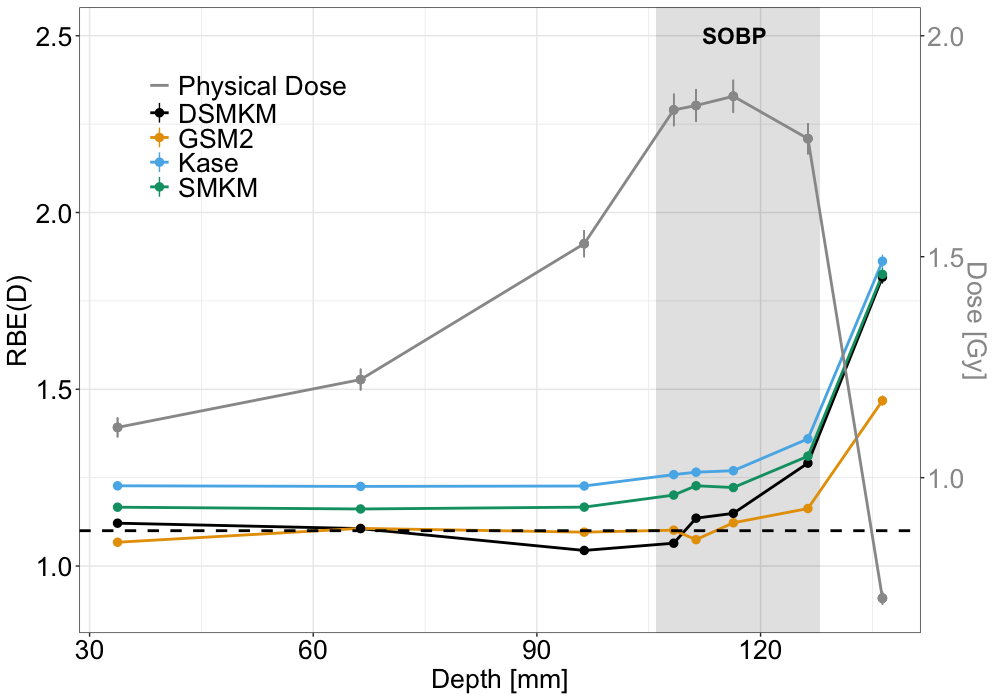}
	\end{minipage}}
	\caption{Radiobiological characterization of a proton SOBP. Cell survival fraction (panel (a)) and dose-dependent RBE (panel (b)), as a function of \red{depth in water}, predicted by DSMKM (black), GSM$^2$ (orange), MKM (light blue), and mSMKM (green). The SOBP region is depicted in a shaded grey. The horizontal dashed black line in panel (b) correspond to RBE 1.1, which is the standard value used in clinics.}
	\label{fig:DoseAndRBE_SOBP}
\end{figure}

Figure \ref{fig:DoseAndRBE_SOBP} panel (a) shows the depth survival curve compared to the physical dose. All models predict a similar trend: the survival is higher at the entrance, around 0.7, and drops to a minimum in the SOBP, where the survival remains constant at around 0.5. At the distal edge of the SOBP, the survival fraction raises again to a similar value as the entrance channel. However, quantitative differences between the considered models are appreciable. The MKM-z* and mSMKM formulations predict cell survival fractions systematically lower than GSM$^2$ in all regions. The DSMKM shows comparable survival fraction values with GSM$^2$ in the entrance up to the mid-SOBP, while in the distal region, GSM$^2$ predicts a slightly higher cell survival. Dose-dependent cell survivals were further used to calculate RBE according to equation \eqref{eq:RBE}. All the models predict an RBE higher than 1 for all depths, Figure \ref{fig:DoseAndRBE_SOBP} panel (b). In addition, all models show a constant RBE trend as a function of penetration depth with a value between 1.1 and 1.2 in the entrance channel and a sharp increase in the distal edge of the field. Table \ref{tab:HSGCell} reports the survival fraction and RBE values, along with statistical errors, at three water depths: entrance (34 mm), middle SOBP (116 mm), and distal edge (136 mm).

\begin{table}[]
\centering
\begin{tabular}{ccc}
\textbf{}                            & \multicolumn{2}{c}{\textbf{Entrance}}  \\ \cline{2-3}
\multicolumn{1}{c}{\textbf{}}        & \textbf{S(D)} & \multicolumn{1}{c}{\textbf{RBE(D)}} \\ \hline
 \textbf{GSM$^2$} & $0.74\pm0.03$ & \multicolumn{1}{c}{$1.06\pm0.05$} \\
\textbf{MKM-z*}  & $0.70\pm0.02$ & \multicolumn{1}{c}{$1.22\pm0.04$} \\
\textbf{mSMKM} & $0.69\pm0.02$ & \multicolumn{1}{c}{$1.16\pm0.04$} \\
\textbf{DSMKM}   & $0.73\pm0.03$  & \multicolumn{1}{c}{$1.12\pm0.05$} \\
\hline
& & \\

\textbf{}                            & \multicolumn{2}{c}{\textbf{Mid-SOBP}}  \\ \cline{2-3}
\multicolumn{1}{c}{\textbf{}}        & \textbf{S(D)} & \multicolumn{1}{c}{\textbf{RBE(D)}} \\ \hline
 \textbf{GSM$^2$} & $0.54\pm0.02$ & \multicolumn{1}{c}{$1.12\pm0.05$} \\
\textbf{MKM-z*}  & $0.48\pm0.01$ & \multicolumn{1}{c}{$1.27\pm0.03$} \\
\textbf{mSMKM} & $0.50\pm0.01$ & \multicolumn{1}{c}{$1.22\pm0.03$} \\
\textbf{DSMKM}   & $0.53\pm0.02$  & \multicolumn{1}{c}{$1.15\pm0.05$} \\
\hline
& & \\

\textbf{}                            & \multicolumn{2}{c}{\textbf{Distal}}  \\ \cline{2-3}
\multicolumn{1}{c}{\textbf{}}        & \textbf{S(D)} & \multicolumn{1}{c}{\textbf{RBE(D)}} \\ \hline
 \textbf{GSM$^2$} & $0.77\pm0.07$ & \multicolumn{1}{c}{$1.47\pm0.08$} \\
\textbf{MKM-z*}  & $0.70\pm0.04$ & \multicolumn{1}{c}{$1.86\pm0.07$} \\
\textbf{mSMKM} & $0.71\pm0.04$ & \multicolumn{1}{c}{$1.82\pm0.07$} \\
\textbf{DSMKM}   & $0.72\pm0.07$  & \multicolumn{1}{c}{$1.81\pm0.09$} \\
\bottomrule
\end{tabular}
\caption{Cell survival fraction, S, and RBE calculated with MKM formulations and GSM$^2$ for HSG cell line. The values are reported at three relevant water depths of the proton SOBP: entrance (34 mm), mid-SOBP (116 mm), and distal penumbra (136 mm).}
\label{tab:HSG&V79Summary}
\end{table}

\subsection{MONAS application to a patient case}\label{sec:Patient}

In order to estimate a 3D spatial distribution of cell survival and RBE for a patient case, we calculated lookup tables (LUTs) of radiobiological parameters $\alpha$ and $\beta$ for protons specific to the HSG cell line.
%Panel (a) shows $\alpha$ values as a function of proton kinetic energy for each model implemented in the MONAS code. All curves follow the same trend: $\alpha$ values sharply decrease from 1.2 $Gy^{-1}$ at 0.1 MeV to around 0.3 $Gy^{-1}$ at 10 MeV. Above this energy, the $\alpha$ variations are less marked. $\beta$ values predicted by the MKM-z* formulation with MONAS are constant as a function of proton kinetic energy around 0.07 and 0.08 $Gy^{-2}$, Figure \ref{fig:AlphaBetaLUT} panel (b). GSM$^2$ model predicts a decreasing trend of $\beta$ from 0.10 $Gy^{-2}$ at 1 MeV to 0.04 $Gy^{-2}$ above 70 MeV. Below 1 MeV proton energy, the mSMKM and DSMKM models show an increasing trend of beta which saturates around 0.08 $Gy^{-2}$ above 1 MeV proton energy.

\begin{comment}
\begin{figure}
 \centering
 \subfloat[]{
	\begin{minipage}[c][1\width]{
	   0.49\textwidth}
	   \centering
	   \includegraphics[width=1\textwidth]{figures/Alpha_H_spline.png}
	\end{minipage}}
	\hfill 
	\subfloat[]{
	\begin{minipage}[c][\width]{
	   0.49\textwidth}
	   \centering
	   %\includegraphics[width=1\textwidth]{figures/RBEd}
	   \includegraphics[width=1\textwidth]{figures/Beta_H_spline.png}
	\end{minipage}}

	\caption{Panel (a) and (b) show the $\alpha$ and $\beta$ values, respectively, for the HSG cell line as a function of proton kinetic energy extrapolated from cell survival curves predicted by MONAS code.}
	\label{fig:AlphaBetaLUT}
\end{figure}
\end{comment}

\begin{figure}
 \centering
 \subfloat[]{
	\begin{minipage}[c][.8\width]{
	   0.49\textwidth}
	   \centering
	   \includegraphics[width=1\textwidth]{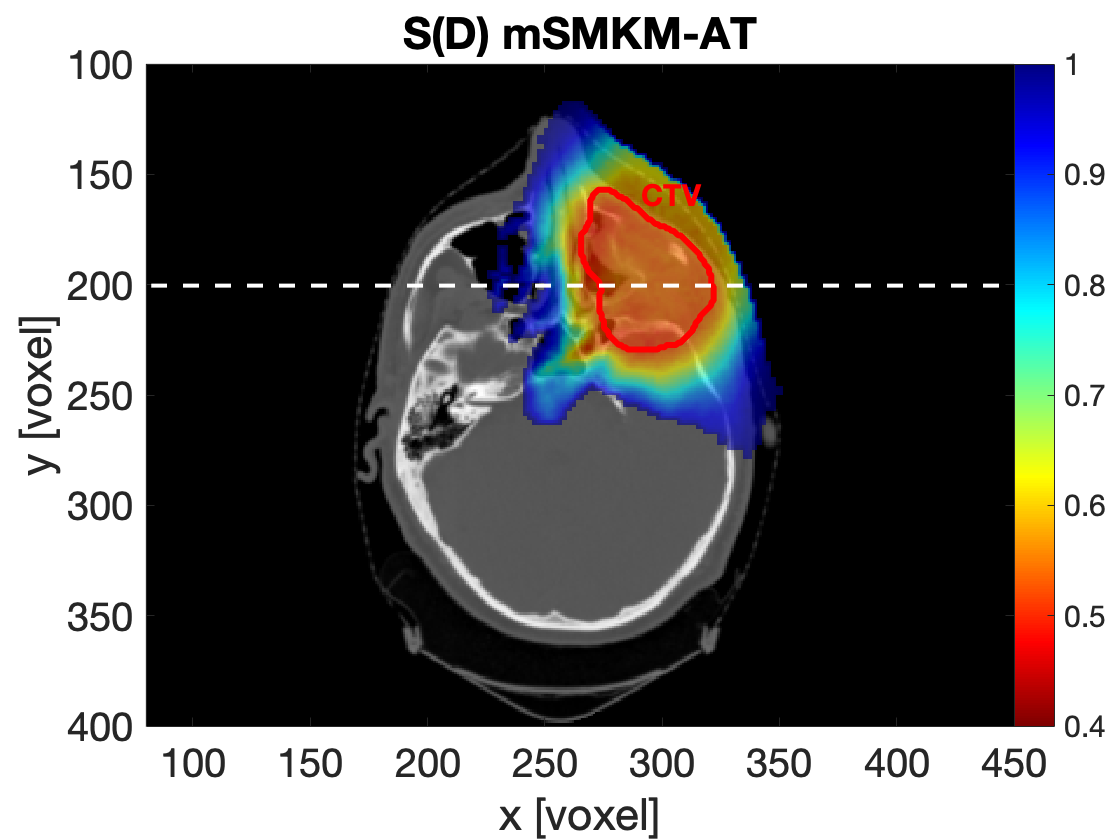}
	\end{minipage}}
	\hfill 
	\subfloat[]{
	\begin{minipage}[c][.8\width]{
	   0.49\textwidth}
	   \centering
	   \includegraphics[width=1\textwidth]{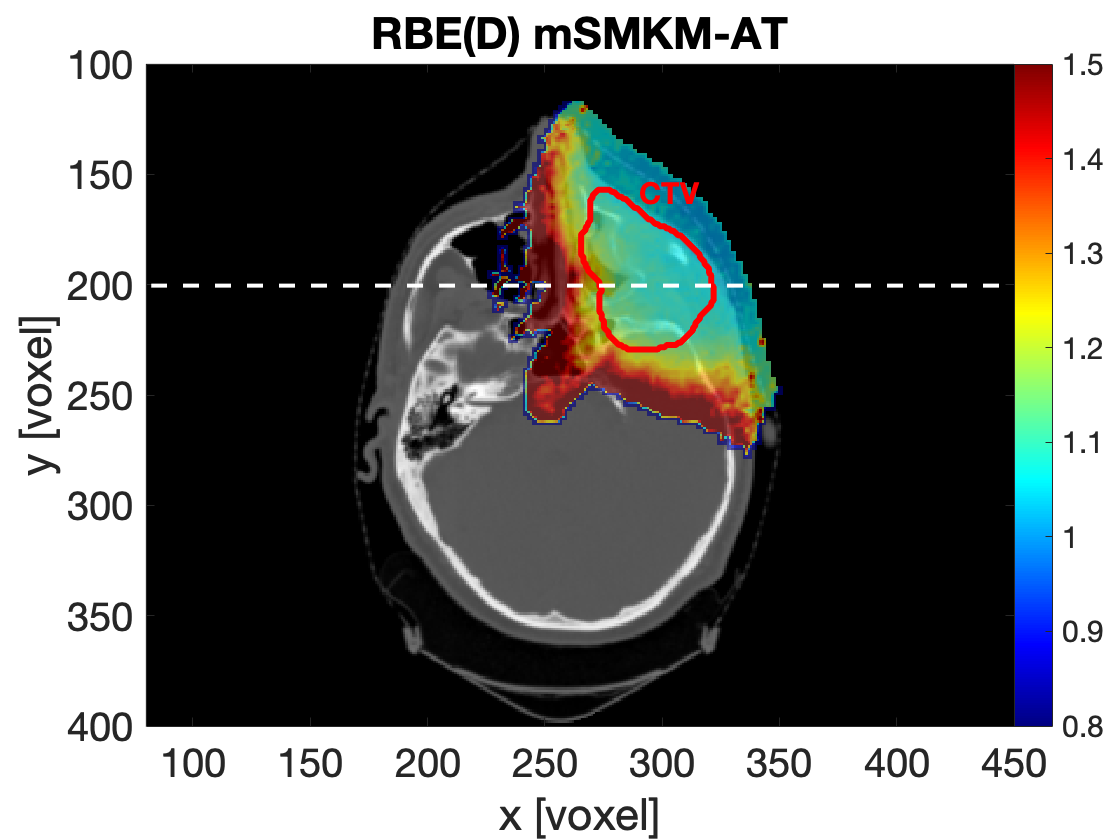}
	\end{minipage}}\\
 \subfloat[]{
	\begin{minipage}[c][.8\width]{
	   0.49\textwidth}
	   \centering
	   \includegraphics[width=1\textwidth]{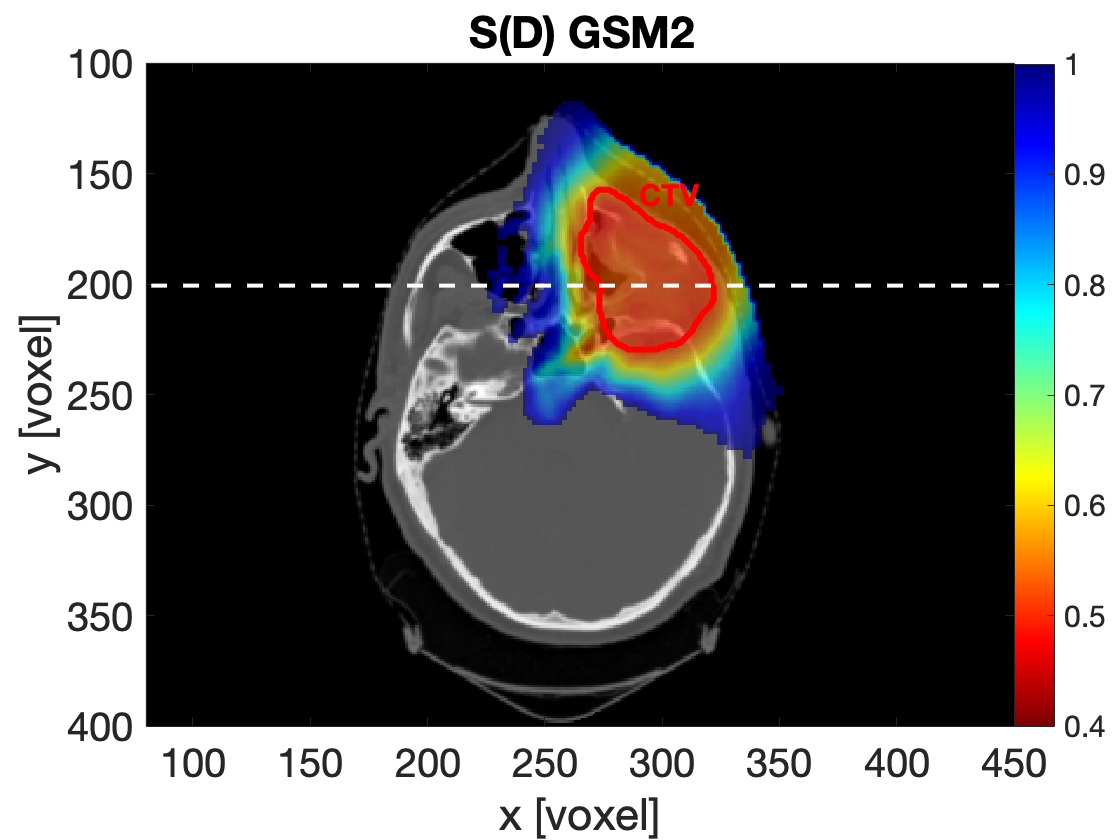}
	\end{minipage}}
	\hfill 
	\subfloat[]{
	\begin{minipage}[c][.8\width]{
	   0.49\textwidth}
	   \centering
	   \includegraphics[width=1\textwidth]{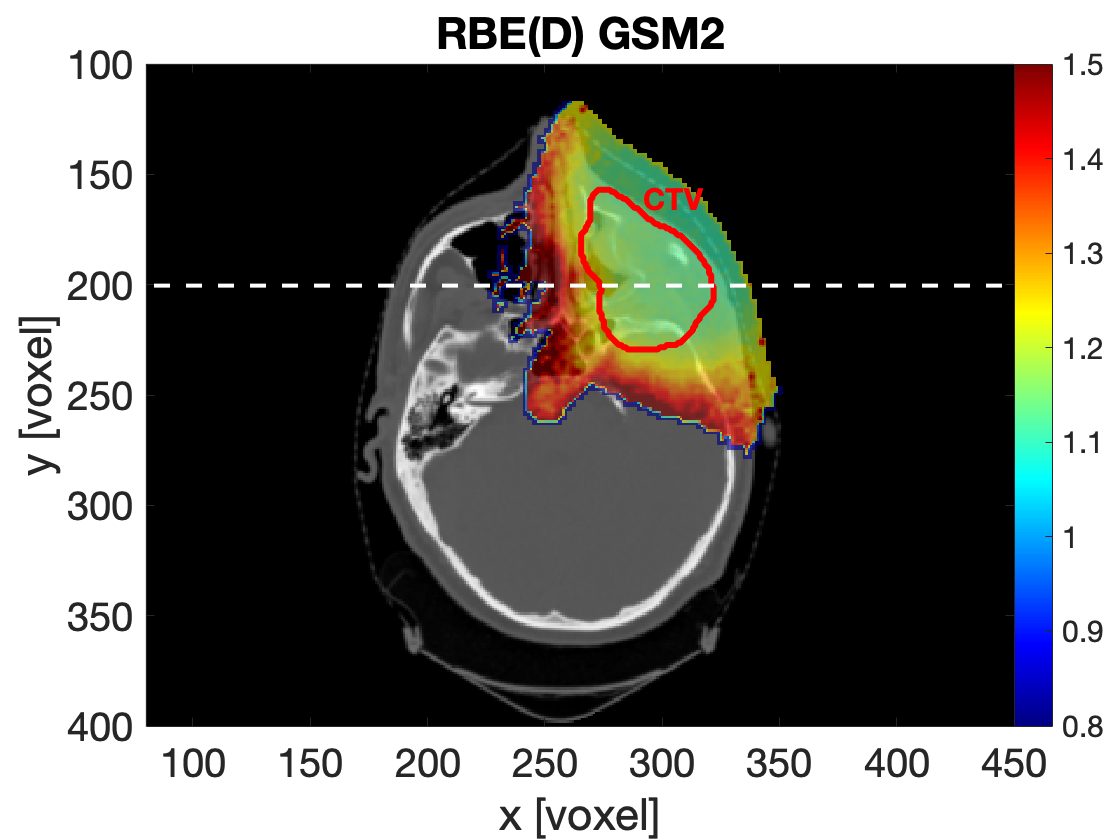}
	\end{minipage}}\\
 \subfloat[]{
	\begin{minipage}[c][.8\width]{
	   0.49\textwidth}
	   \centering
	   \includegraphics[width=1\textwidth]{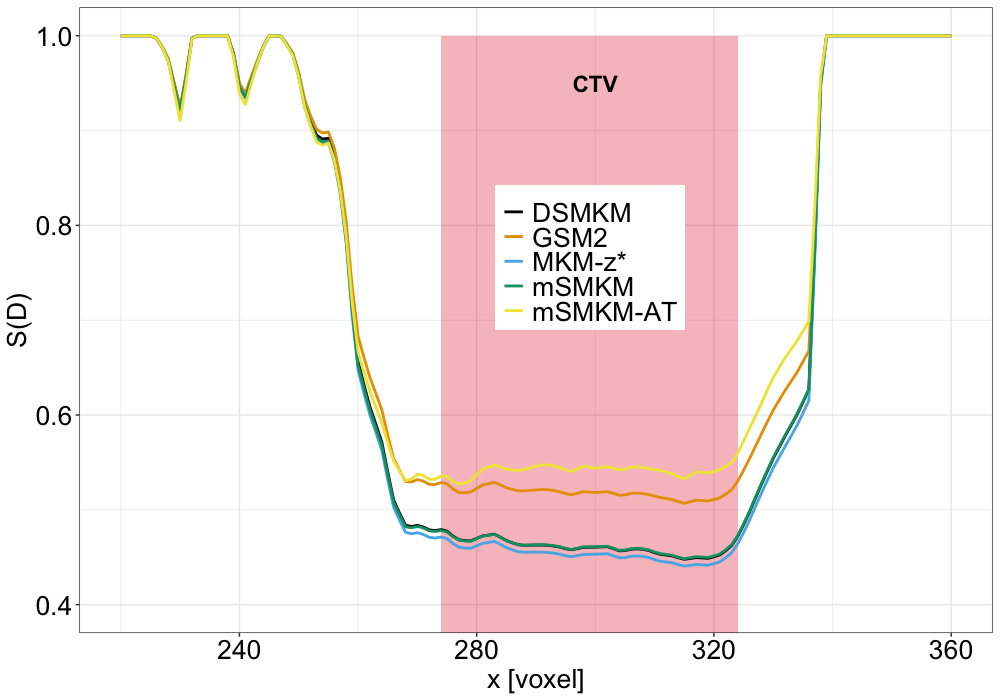}
	\end{minipage}}
	\hfill 
	\subfloat[]{
	\begin{minipage}[c][.8\width]{
	   0.49\textwidth}
	   \centering
	   \includegraphics[width=1\textwidth]{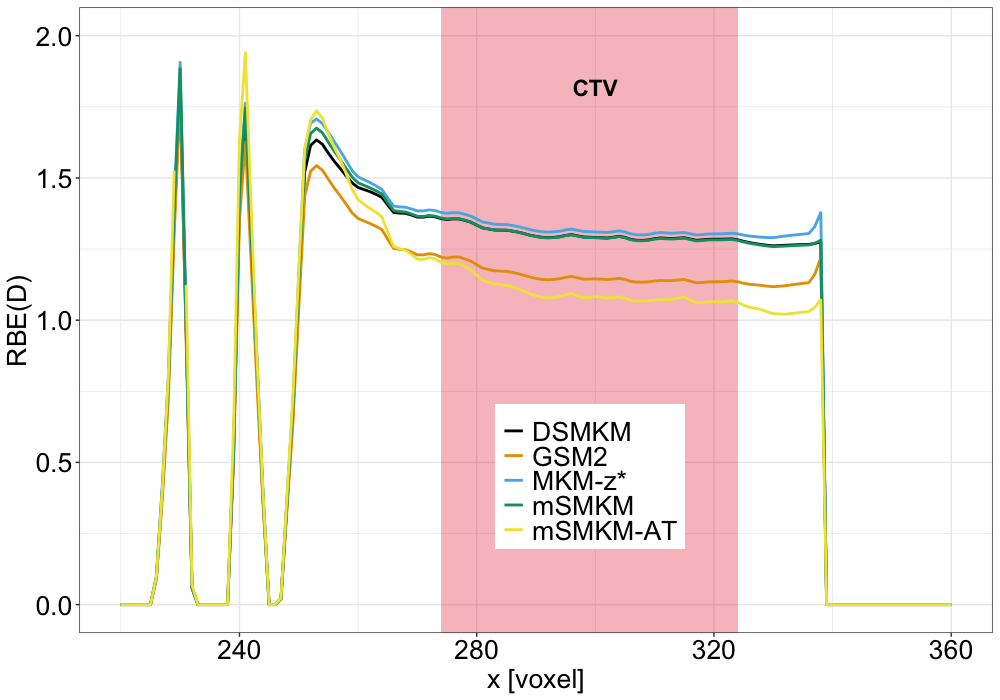}
	\end{minipage}}\\
	\caption{Panels (a) and (b) show the 2D spatial distributions of cell survival and RBE calculated with mSMKM-AT on the patient case, respectively. The same patient slice and distributions are shown for the GSM2 model, representative of the \red{MONAS} prediction (panels (c) and (d)). The horizontal dotted white line indicates the position of the line curves plotted in panels (e) and (f) for cell survival and RBEs, respectively. The red vertical band in the latter plots shows the position of the target region.}
	\label{fig:Patient}
\end{figure}

We incorporated the LUTs within the TOPAS Monte Carlo algorithm and simulated a treatment plan, calculating the 3D spatial distributions of dose-dependent cell survival and RBE. To benchmark our results, we also included the mSMKM with amorphous track (mSMKM-AT), a validated radiobiological model for Carbon ion therapy, \cite{inaniwa2018adaptation}. The 2D spatial distribution of cell survival fraction is plotted in Figure \ref{fig:Patient} for GSM$^2$ (panel (c)), and mSMKM-AT (panel (a)) as a benchmark to the current state-of-the-art of MKM clinical application (tuned for carbon ion therapy). The two models agree in the shape of $S(D)$ distribution where a minimum (color map hot region) of the cell survival fraction is estimated in the Clinical Target Volume (CTV) and a few millimeters in the surrounding volume. By comparing the \red{monodimensional} depth-survival curves (panel (e)), large differences are evident among the models. MONAS-based MKM formulations predict the lowest $S(D)$ values (0.47 $\pm$ 0.03 on average in CTV), while GSM$^2$ and mSMKM-AT predict an average cell survival fraction of 0.51$\pm$ 0.07 and 0.55 $\pm$ 0.04, respectively.

The RBE 2D map displayed in panels (b) and (d) exhibits similar patterns: both the mSMKM-AT and GSM$^2$ are characterized by a notable increase in RBE values in the distal region of the therapeutic field, specifically at beam angles of 30 and 60 degrees. mSMKM-AT 2D distribution shows a wider dark red region (RBE values above 1.5) compared to GSM$^2$ in the distal part of the CTV, while in the patient's entrance mSMKM-AT map has lower RBE values (blue colormap). As shown in the depth-RBE curve (panel (f)), MKM models from MONAS code predict RBE values systematically higher than GSM$^2$ and mSMKM-AT both at the beam entrance and in the distal region (from 1.3 to 1.6). GSM$^2$ model predicts RBE values lower than the MKM formulations from around 1.1 at the entrance, up to 1.5 downstream of the CTV. The same holds for mSMKM-AT which shows RBE values around 1 at the entrance, exceeding GSM$^2$ in distal reaching RBEs of around 1.7.

Although the inter-model differences in absolute values of the survival $S$ and RBE, both MONAS-based and mSMKM-AT models show the same spatial distribution characteristics. Inside the target volume $S$ and RBEs remain constant and sharply increase at the distal edges of the fields where most of the organs at risk are located.

\section{Discussion}

In this work, we presented the MONAS toolkit, a TOPAS MC extension that combines accurate Monte Carlo simulations of microdosimetric distributions with microdosimetry-based models for predicting cell survival and RBE. %MONAS extends the previous microdosimetric extension \cite{zhu2019microdosimetric}.
MONAS provides a powerful tool to bridge the gap between a microdosimetric description of a radiation field, the biological evaluation of radiation effects, and its implementation in the clinical practice of proton beam therapy. 

To show MONAS applications, we presented two examples: i) the radiobiological characterization of a SOBP, and ii) the calculation of the RBE spatial distribution for a real patient's treatment with protons. In Figure \ref{fig:DoseAndRBE_SOBP}, we showed that MONAS can estimate the cell survival fraction and RBEs from experimentally validated spectra, predicting a trend consistent with previous works \cite{kase2006microdosimetric, tran2017characterization}. 
In this context, it is important to emphasize that most of the research paper on protons and heavier ions focuses on RBE$_{10}$, which is the ratio of reference radiation and ion dose giving the 10$\%$ survival fraction \cite{bianchi2020microdosimetry, conte2020microdosimetry, debrot2018soi, tran2017characterization}. The RBE$_{10}$ represents a universally accepted radiobiological endpoint, allowing thus a comparison of different radiobiological experiments. However, when considering clinical applications, it becomes more meaningful to examine a dose-dependent RBE, i.e. the RBE associated with the dose specifically absorbed by each voxel of the treatment plan. This critical information enables direct estimation of the treatment plan's biological effectiveness. The presented MONAS extension, which is directly linked to TOPAS, offers a rapid assessment of the MC-based dose-dependent RBE, which can have an impact both in clinical scenarios and in experimental radiobiology.

Using the complete range of microdosimetric spectrum as the input for radiobiological models emphasized the need to recalibrate the model parameters, as outlined in Table \ref{tab:parameters}, in order to accurately fit the MONAS cell survival curves with the existing experimental data. Figure \ref{fig:HSGSurvival} shows a good agreement between MONAS predictions and the experimental \textit{in vitro} data at different irradiation conditions. Furthermore, the new set of parameters is consistent with parameters already published for the MKM \cite{kase2006microdosimetric, sato2012cell, inaniwa2018adaptation} in terms of the order of magnitude of the cell domain and nucleus radii, as well as the $\alpha$-$\beta$ absolute values. 
%The need to re-calibrate the free parameters of all radiobiological models is related to the description of the energy deposition. While the recent formulation of the MKM relies on the amorphous track structure model, MONAS employs microdosimetry. 
%due to the change in the characterization of particle energy deposition at the micrometer scale by the Monte Carlo algorithm, compared to the amorphous track structure model used by the recent formulations of MKM models.
%The results reported indicate that all radiobiological models heavily depend on the physical description of the radiation field used for assessing the biological outcome. 
The MONAS toolkit thus further allows testing the robustness of the model parameter against experiments. In MONAS, all microdosimetric values are evaluated using the entire microdosimetric distribution (Figures \ref{fig:SpecifiEnergySpectraDomain}-\ref{fig:SpecifiEnergySpectraCell}), which can be validated by direct comparison with the experimental $y$-spectra measured with commercial microdosimetric detectors. As a result, the MONAS toolkit provides a comprehensive framework for validating the entire radiobiological workflow, encompassing the simulation of the radiation field's physical properties and the direct estimation of its biological impact. This is achieved by utilizing microdosimetric quantities and cell survival experiments that are directly calibrated against experimental data, ensuring accuracy and reliability in assessing biological effectiveness.

The second main application of MONAS is the RBE evaluation in a real patient case. In Figure \ref{fig:Patient}, we show the 3D spatial distribution of cell survival fraction and RBE values calculated for proton treatment of a head and neck tumor. The results for a complex beam and patient's geometry follow the same trend exhibited in Figure \ref{fig:DoseAndRBE_SOBP} from the SOBP: RBE hot spots are observed at the distal edge of the field, where most of the organs at risk are located (e.g. salivary glands, oral cavity, optical nerves). We used the MONAS extension to generate LUTs of $\alpha$ and $\beta$ coefficients for monoenergetic proton beams and eventually asses the biological effectiveness of a mixed-energy beam averaging $\alpha$ and $\beta$ values. 
%Figure \ref{fig:AlphaBetaLUT}, shows an overall good agreement between the models for the $\alpha$ parameters, while the $\beta$ parameter shows a significantly different trend, with however a common flattening of the $\beta$ value at high energies. Different predictions on the $\beta$ values have already been reported in the literature when comparing radiobiological models \cite{bellinzona2021linking}. In particular, the MKM predicts a constant $\beta$ for different LET radiation, which is contradicted by experimental cell survival data. To overcome this limitation, and include a variable $\beta$, the DSMKM and the mSMKM have been developed (Figure \ref{fig:AlphaBetaLUT}). GSM$^2$ cell survival prediction depends on the whole microdosimetric spectrum, \cite{cordoni2022cell}, with both the linear and quadratic terms depending on radiation quality. This is translated into a more complex dependence of the $\beta$ parameter with respect to the LET. 
The RBE evaluated in the present paper is calculated considering only the contribution from primary and secondary protons, but it can be extended to include other secondary ions by adding LUTs for alpha, carbons, and oxygen ions. This could increase the accuracy of the estimated RBE values, especially at the field-edge and out-of-field, where a non-negligible contribution to the radiation field is given by high-LET radiation such as alpha particles \cite{missiaggia2023investigation}.
The application of MONAS in a real patient case represents a significant advancement in the integration of a detailed microdosimetric description of the radiation field into treatment planning systems. These LUTs can be incorporated into a particle tracking Monte Carlo toolkit to estimate a three-dimensional map of cell survival fraction and RBE within a mixed-field radiation treatment plan. 

While the use of LUTs does not fully address the computational challenge of simulating the complete microdosimetric distribution in all voxels, we propose a fundamental step into the usage of microdosimetric information in proton and ion therapy. LUTs have been generated by simulating the entire microdosimetry distribution of monoenergetic beams, which were then used as input for the radiobiological models to estimate the cell survival fraction. Therefore, the intrinsic stochastic nature of energy deposition at the micron scale is taken into account in the final LUT values. The use of pre-calculated $\alpha$ and $\beta$ values that are loaded during the particle simulation in patients does not increase the treatment plan computation time. In fact, the time required by TOPAS MC to calculate $\alpha_{mix}$ and $\sqrt{\beta_{mix}}$ in each voxel of the patient is comparable to a similar volumetric scorer already implemented in TOPAS (e.g. \textit{ProtonLET} \cite{cortes2015critical, granville2015comparison}). This feature is of great significance in the proposed methodology because it enables the inclusion of our LUTs in clinical Monte Carlo treatment planning systems without causing a slowdown in computational time for treatment plan calculations.

Our findings reveal that there is a clear inter-model variability in the absolute values of cell survival and, thus in RBE values observed both in SOBP characterization and in the patient simulation, as seen in Figures \ref{fig:DoseAndRBE_SOBP}-\ref{fig:Patient}. Figures \ref{fig:Patient} panels (e)-(f) exhibit significant differences between the MKM models and GSM$^2$ in the CTV region, being further the mSMKM-AT closer to the GSM$^2$ prediction. Differences in the models stem from the fact that they all have different foundations and include different stochasticities. In addition, deviations in the predictions of RBE have already been verified in the literature when comparing the MKM with the LEM, \cite{monini2019comparison,bellinzona2021linking}. Furthermore, the predictions of the implemented models show similar patterns, with GSM$^2$ predicting a higher survival probability in the tumor compared to all MKM versions. The aim of this study is to provide a proof-of-principle of  MONAS key features without making any assumptions about the model's parameters. The strength of our toolkit lies in the ability to customize each model's parameters and cell line-specific settings. However, we provided a set of potential parameters that yield radiobiological results consistent with the \textit{in vitro} data presented in Figure \ref{fig:HSGSurvival}.
Further investigations will be conducted to calculate the model parameters that best fit a larger sample of experimental cell survival curves at various irradiation conditions. This analysis highlights the importance of determining precise model parameters that have a significant impact on the absolute values of cell survival fractions and, consequently, RBE. Furthermore, all models predict an RBE that significantly differs from the constant value used in the clinic. GSM$^2$ and the DSMKM calculate an RBE value close to 1.1 only in the entrance channel, while all the models show a sharp increase in the RBE in the distal region. This increase in the RBE could have an impact on the organs at risk located right after the tumor, for which a significant underestimation of RBE could lead to toxicities.

Microdosimetry is proving to be extremely valuable in clinical applications for several compelling reasons, surpassing the traditional use of LET values and providing a more reliable and experimentally measurable, as demonstrated in numerous campaigns conducted over the years, \cite{kase2006microdosimetric, tran2018relative,missiaggia2020microdosimetric,bianchi2020microdosimetry, lee2021estimating,missiaggia2023investigation, magrin2023state}, radiation quality description. Moreover, microdosimetry naturally includes the geometry of the sensitive volume in the measured spectra, providing information that not only considers the stochastic nature of energy deposition but also incorporates specific geometric considerations of the studied volume. This eliminates any potential uncertainty in interpreting the analyzed spectra and quantities. LET often fails to specify whether the value represents the track average or dose average. Additionally, different LET scorers are commonly used, and including only primary particles or both primary and secondary particles in the scorer can completely change the physical significance of the resulting values and the estimation of the biological effect \cite{grassberger2011elevated, bellinzona2021biological, kalholm2021systematic}. This becomes particularly problematic in the out-of-field regions, where the radiation field is mixed and short-track particles can play a significant role, and LET is known to be a bad descriptor of the radiation quality, \cite{grun2019dose}. On the other hand, microdosimetry offers comprehensive information that allows for a precise assessment of all the physical processes involved, enabling a more accurate estimation of the biological effects.

Therefore, radiotherapeutic clinical practice can greatly benefit from consistently and rigorously evaluating treatment plans based on microdosimetry. In the past, the computational effort required for calculating microdosimetry quantities in treatment plans has been a hindrance to its practical implementation in the clinic. However, with advancements in computational power and the effective simplification methods demonstrated in this study, clinicians can now improve the prescribed treatment plans by incorporating microdosimetric considerations allowing for a more robust estimation of the plan's biological effectiveness.

\label{Sec:Res}
\section{Conclusions}
In this work, we presented a novel TOPAS MC extension, MONAS: MicrOdosimetry-based modelliNg for RBE ASsessment, which allows the user to evaluate dose-dependent cell survival curves and RBE with the most used microdosimetry-based radiobiological models: three MKM formulations (saturation corrected MKM (MKM-z*) \cite{kase2006microdosimetric}, Double Stochastic MKM (DSMKM) \cite{sato2012cell} and the modified Stochastic MKM (mSMKM) \cite{inaniwa2018adaptation}) and the GSM$^2$ model, \cite{cordoni2021generalized,cordoni2022cell}.

MONAS wraps the already published TOPAS microdosimetric extensions to evaluate the single- and multi-event specific energy ($z$) distributions at different \red{micrometric} scales. Full microdosimetric distributions are then used as input for both MKM and GSM$^2$ models. This approach showed intrinsic differences in microdosimetric radiation characterization with respect to the amorphous track structure model used in the latest MKM formulations. Therefore, we recalculated the model parameter which best fit the radiobiological experiments for the HSG cell line.
To show the main MONAS applications, we reproduced experimental microdosimetric spectra from a passively scattered SOBP. We used the MONAS code to assess cell survival fraction and RBE as a function of proton penetration depth. Our findings are consistent with the well-known RBE trend which \red{presents} a steep increase in the distal edge of the field. Furthermore, we were able to assess the high inter-model variability on the absolute RBE values\red{thus quantifying a radiobiological uncertainty in proton plans in addition to other physical uncertainties}.

The applicability of the MONAS toolkit can be further extended to a radiobiological analysis of treatment plans. We showed that it is possible to generate radiobiological parameter look-up tables which can be combined with the Monte Carlo toolkit for computing RBE maps on patients and therapeutic beam geometries. We showed an example of cell survival and RBE predictions on a real head and neck proton therapy plan delivered at the Dwoskin Proton therapy center at the University of Miami. The results have been compared to the mSMKM model based on the amorphous track structure model, as recently developed in \cite{inaniwa2018adaptation}, which represents one of the most used models in carbon ion therapy. Despite the variability in RBE absolute values, \red{all} the models showed a reasonable RBE trend as a function of beam penetration depth. Further investigation will focus on a full a biological robust optimization, based on the above mentioned uncertainty. 

In conclusion, the MONAS extension offers a comprehensive microdosimetric framework for assessing the biological effect of radiation in both research and clinical environments. MONAS could be a key tool to include a detailed microdosimetric description of radiation \red{field} into treatment planning systems for variable RBE calculations.
\label{Sec:Con}

%\ack

\cleardoublepage
%\nocite{*}
\bibliographystyle{apalike}
\bibliography{MONAS}

\begin{thebibliography}{}

\bibitem[Agostinelli et~al., 2003]{agostinelli2003geant4}
Agostinelli, S., Allison, J., Amako, K.~a., Apostolakis, J., Araujo, H., Arce,
  P., Asai, M., Axen, D., Banerjee, S., Barrand, G., et~al. (2003).
\newblock Geant4—a simulation toolkit.
\newblock {\em Nuclear instruments and methods in physics research section A:
  Accelerators, Spectrometers, Detectors and Associated Equipment},
  506(3):250--303.

\bibitem[Attili et~al., 2022]{attili2022modelling}
Attili, A., Scifoni, E., and Tommasino, F. (2022).
\newblock Modelling the hprt-gene mutation induction of particle beams:
  systematic in vitro data collection, analysis and microdosimetric kinetic
  model implementation.
\newblock {\em Physics in Medicine \& Biology}, 67(19):195001.

\bibitem[Baratto-Rold{\'a}n et~al., 2021]{baratto2021microdosimetry}
Baratto-Rold{\'a}n, A., Bertolet, A., Baiocco, G., Carabe, A., and
  Cort{\'e}s-Giraldo, M.~A. (2021).
\newblock Microdosimetry and dose-averaged let calculations of protons in
  liquid water: a novel geant4-dna application.
\newblock {\em Frontiers in Physics}, page 631.

\bibitem[Bellinzona et~al., 2021a]{bellinzona2021biological}
Bellinzona, E.~V., Grzanka, L., Attili, A., Tommasino, F., Friedrich, T.,
  Kr{\"a}mer, M., Scholz, M., Battistoni, G., Embriaco, A., Chiappara, D.,
  et~al. (2021a).
\newblock Biological impact of target fragments on proton treatment plans: An
  analysis based on the current cross-section data and a full mixed field
  approach.
\newblock {\em Cancers}, 13(19):4768.

\bibitem[Bellinzona et~al., 2021b]{bellinzona2021linking}
Bellinzona, V., Cordoni, F., Missiaggia, M., Tommasino, F., Scifoni, E.,
  La~Tessa, C., and Attili, A. (2021b).
\newblock Linking microdosimetric measurements to biological effectiveness in
  ion beam therapy: A review of theoretical aspects of mkm and other models.
\newblock {\em Frontiers in Physics}, 8:578492.

\bibitem[Bianchi et~al., 2020]{bianchi2020microdosimetry}
Bianchi, A., Selva, A., Colautti, P., Bortot, D., Mazzucconi, D., Pola, A.,
  Agosteo, S., Petringa, G., Cirrone, G., Reniers, B., et~al. (2020).
\newblock Microdosimetry with a sealed mini-tepc and a silicon telescope at a
  clinical proton sobp of catana.
\newblock {\em Radiation Physics and Chemistry}, 171:108730.

\bibitem[Bianchi et~al., 2022]{bianchi2022topas}
Bianchi, A., Selva, A., Reniers, B., Vanhavere, F., and Conte, V. (2022).
\newblock Topas simulations of the response of a mini-tepc: benchmark with
  experimental data.
\newblock {\em Physics in Medicine and Biology}.

\bibitem[Bradley et~al., 2001]{bradley2001solid}
Bradley, P., Rosenfeld, A., and Zaider, M. (2001).
\newblock Solid state microdosimetry.
\newblock {\em Nuclear Instruments and Methods in Physics Research Section B:
  Beam Interactions with Materials and Atoms}, 184(1-2):135--157.

\bibitem[Cartechini, 2023]{github}
Cartechini, G. (2023).
\newblock Monas gitlab repository.

\bibitem[Chatterjee and Schaefer, 1976]{chatterjee1976microdosimetric}
Chatterjee, A. and Schaefer, H. (1976).
\newblock Microdosimetric structure of heavy ion tracks in tissue.
\newblock {\em Radiation and environmental biophysics}, 13(3):215--227.

\bibitem[Conte et~al., 2020]{conte2020microdosimetry}
Conte, V., Agosteo, S., Bianchi, A., Bolst, D., Bortot, D., Catalano, R.,
  Cirrone, G., Colautti, P., Cuttone, G., Guatelli, S., et~al. (2020).
\newblock Microdosimetry of a therapeutic proton beam with a mini-tepc and a
  microplus-bridge detector for rbe assessment.
\newblock {\em Physics in Medicine \& Biology}, 65(24):245018.

\bibitem[Cordoni et~al., 2021]{cordoni2021generalized}
Cordoni, F., Missiaggia, M., Attili, A., Welford, S., Scifoni, E., and
  La~Tessa, C. (2021).
\newblock Generalized stochastic microdosimetric model: The main formulation.
\newblock {\em Physical Review E}, 103(1):012412.

\bibitem[Cordoni et~al., 2022a]{cordoni2022multiple}
Cordoni, F.~G., Missiaggia, M., La~Tessa, C., and Scifoni, E. (2022a).
\newblock Multiple levels of stochasticity accounted for in different radiation
  biophysical models: from physics to biology.
\newblock {\em International Journal of Radiation Biology}, pages 1--16.

\bibitem[Cordoni et~al., 2022b]{cordoni2022cell}
Cordoni, F.~G., Missiaggia, M., Scifoni, E., and La~Tessa, C. (2022b).
\newblock Cell survival computation via the generalized stochastic
  microdosimetric model (gsm2); part i: The theoretical framework.
\newblock {\em Radiation Research}, 197(3):218--232.

\bibitem[Cort{\'e}s-Giraldo and Carabe, 2015]{cortes2015critical}
Cort{\'e}s-Giraldo, M. and Carabe, A. (2015).
\newblock A critical study of different monte carlo scoring methods of dose
  average linear-energy-transfer maps calculated in voxelized geometries
  irradiated with clinical proton beams.
\newblock {\em Physics in Medicine \& Biology}, 60(7):2645.

\bibitem[De~Nardo et~al., 2004]{de2004mini}
De~Nardo, L., Cesari, V., Don{\`a}, G., Magrin, G., Colautti, P., Conte, V.,
  and Tornielli, G. (2004).
\newblock Mini-tepcs for radiation therapy.
\newblock {\em Radiation protection dosimetry}, 108(4):345--352.

\bibitem[Debrot et~al., 2018]{debrot2018soi}
Debrot, E., Tran, L., Chartier, L., Bolst, D., Guatelli, S., Vandevoorde, C.,
  de~Kock, E., Beukes, P., Symons, J., Nieto-Camero, J., et~al. (2018).
\newblock Soi microdosimetry and modified mkm for evaluation of relative
  biological effectiveness for a passive proton therapy radiation field.
\newblock {\em Physics in Medicine \& Biology}, 63(23):235007.

\bibitem[Durante et~al., 2017]{durante2017charged}
Durante, M., Orecchia, R., and Loeffler, J.~S. (2017).
\newblock Charged-particle therapy in cancer: clinical uses and future
  perspectives.
\newblock {\em Nature Reviews Clinical Oncology}, 14(8):483--495.

\bibitem[Friedrich et~al., 2013]{friedrich2013systematic}
Friedrich, T., Scholz, U., Els{\"a}Sser, T., Durante, M., and Scholz, M.
  (2013).
\newblock Systematic analysis of rbe and related quantities using a database of
  cell survival experiments with ion beam irradiation.
\newblock {\em Journal of radiation research}, 54(3):494--514.

\bibitem[Furusawa et~al., 2000]{furusawa2000inactivation}
Furusawa, Y., Fukutsu, K., Aoki, M., Itsukaichi, H., Eguchi-Kasai, K., Ohara,
  H., Yatagai, F., Kanai, T., and Ando, K. (2000).
\newblock Inactivation of aerobic and hypoxic cells from three different cell
  lines by accelerated 3he-, 12c-and 20ne-ion beams.
\newblock {\em Radiation research}, 154(5):485--496.

\bibitem[Granville and Sawakuchi, 2015]{granville2015comparison}
Granville, D.~A. and Sawakuchi, G.~O. (2015).
\newblock Comparison of linear energy transfer scoring techniques in monte
  carlo simulations of proton beams.
\newblock {\em Physics in Medicine \& Biology}, 60(14):N283.

\bibitem[Grassberger and Paganetti, 2011]{grassberger2011elevated}
Grassberger, C. and Paganetti, H. (2011).
\newblock Elevated let components in clinical proton beams.
\newblock {\em Physics in Medicine \& Biology}, 56(20):6677.

\bibitem[Gr{\"u}n et~al., 2019]{grun2019dose}
Gr{\"u}n, R., Friedrich, T., Traneus, E., and Scholz, M. (2019).
\newblock Is the dose-averaged let a reliable predictor for the relative
  biological effectiveness?
\newblock {\em Medical physics}, 46(2):1064--1074.

\bibitem[Hawkins, 1996]{hawkins1996microdosimetric}
Hawkins, R. (1996).
\newblock A microdosimetric-kinetic model of cell death from exposure to
  ionizing radiation of any let, with experimental and clinical applications.
\newblock {\em International journal of radiation biology}, 69(6):739--755.

\bibitem[Hawkins, 1994]{hawkins1994statistical}
Hawkins, R.~B. (1994).
\newblock A statistical theory of cell killing by radiation of varying linear
  energy transfer.
\newblock {\em Radiation research}, 140(3):366--374.

\bibitem[Hawkins, 2003]{hawkins2003microdosimetric}
Hawkins, R.~B. (2003).
\newblock A microdosimetric-kinetic model for the effect of non-poisson
  distribution of lethal lesions on the variation of rbe with let.
\newblock {\em Radiation research}, 160(1):61--69.

\bibitem[Inaniwa et~al., 2010]{inaniwa2010treatment}
Inaniwa, T., Furukawa, T., Kase, Y., Matsufuji, N., Toshito, T., Matsumoto, Y.,
  Furusawa, Y., and Noda, K. (2010).
\newblock Treatment planning for a scanned carbon beam with a modified
  microdosimetric kinetic model.
\newblock {\em Physics in Medicine \& Biology}, 55(22):6721.

\bibitem[Inaniwa and Kanematsu, 2018]{inaniwa2018adaptation}
Inaniwa, T. and Kanematsu, N. (2018).
\newblock Adaptation of stochastic microdosimetric kinetic model for
  charged-particle therapy treatment planning.
\newblock {\em Physics in Medicine \& Biology}, 63(9):095011.

\bibitem[Inaniwa et~al., 2013]{inaniwa2013effects}
Inaniwa, T., Suzuki, M., Furukawa, T., Kase, Y., Kanematsu, N., Shirai, T., and
  Hawkins, R.~B. (2013).
\newblock Effects of dose-delivery time structure on biological effectiveness
  for therapeutic carbon-ion beams evaluated with microdosimetric kinetic
  model.
\newblock {\em Radiation research}, 180(1):44--59.

\bibitem[J{\"a}kel et~al., 2016]{jakel2016icru}
J{\"a}kel, O., Bert, C., Fossati, P., and Kamada, T. (2016).
\newblock Icru report 93: prescribing, recording, and reporting light ion beam
  therapy.
\newblock {\em J ICRU}, 16(1-2):37--58.

\bibitem[Jarlskog and Paganetti, 2008]{jarlskog2008physics}
Jarlskog, C.~Z. and Paganetti, H. (2008).
\newblock Physics settings for using the geant4 toolkit in proton therapy.
\newblock {\em IEEE Transactions on nuclear science}, 55(3):1018--1025.

\bibitem[Kalholm et~al., 2021]{kalholm2021systematic}
Kalholm, F., Grzanka, L., Traneus, E., and Bassler, N. (2021).
\newblock A systematic review on the usage of averaged let in radiation biology
  for particle therapy.
\newblock {\em Radiotherapy and Oncology}, 161:211--221.

\bibitem[Kase et~al., 2007]{kase2007biophysical}
Kase, Y., Kanai, T., Matsufuji, N., Furusawa, Y., Els{\"a}sser, T., and Scholz,
  M. (2007).
\newblock Biophysical calculation of cell survival probabilities using
  amorphous track structure models for heavy-ion irradiation.
\newblock {\em Physics in Medicine \& Biology}, 53(1):37.

\bibitem[Kase et~al., 2006]{kase2006microdosimetric}
Kase, Y., Kanai, T., Matsumoto, Y., Furusawa, Y., Okamoto, H., Asaba, T.,
  Sakama, M., and Shinoda, H. (2006).
\newblock Microdosimetric measurements and estimation of human cell survival
  for heavy-ion beams.
\newblock {\em Radiation research}, 166(4):629--638.

\bibitem[Kiefer and Straaten, 1986]{kiefer1986model}
Kiefer, J. and Straaten, H. (1986).
\newblock A model of ion track structure based on classical collision dynamics
  (radiobiology application).
\newblock {\em Physics in Medicine \& Biology}, 31(11):1201.

\bibitem[Lee et~al., 2021]{lee2021estimating}
Lee, S.~H., Mizushima, K., Kohno, R., Iwata, Y., Yonai, S., Shirai, T., Pan,
  V.~A., Bolst, D., Tran, L.~T., Rosenfeld, A.~B., et~al. (2021).
\newblock Estimating the biological effects of helium, carbon, oxygen, and neon
  ion beams using 3d silicon microdosimeters.
\newblock {\em Physics in Medicine \& Biology}, 66(4):045017.

\bibitem[Loeffler and Durante, 2013]{loeffler2013charged}
Loeffler, J.~S. and Durante, M. (2013).
\newblock Charged particle therapy—optimization, challenges and future
  directions.
\newblock {\em Nature reviews Clinical oncology}, 10(7):411--424.

\bibitem[Magrin et~al., 2023]{magrin2023state}
Magrin, G., Palmans, H., Stock, M., and Georg, D. (2023).
\newblock State-of-the-art and potential of experimental microdosimetry in
  ion-beam therapy.
\newblock {\em Radiotherapy and Oncology}, page 109586.

\bibitem[Mairani et~al., 2022]{mairani2022roadmap}
Mairani, A., Mein, S., Blakely, E., Debus, J., Durante, M., Ferrari, A., Fuchs,
  H., Georg, D., Grosshans, D.~R., Guan, F., et~al. (2022).
\newblock Roadmap: helium ion therapy.
\newblock {\em Physics in Medicine \& Biology}, 67(15):15TR02.

\bibitem[Manganaro et~al., 2018]{manganaro2018survival}
Manganaro, L., Russo, G., Bourhaleb, F., Fausti, F., Giordanengo, S., Monaco,
  V., Sacchi, R., Vignati, A., Cirio, R., and Attili, A. (2018).
\newblock ‘survival’: a simulation toolkit introducing a modular approach
  for radiobiological evaluations in ion beam therapy.
\newblock {\em Physics in Medicine \& Biology}, 63(8):08NT01.

\bibitem[Manganaro et~al., 2017]{manganaro2017monte}
Manganaro, L., Russo, G., Cirio, R., Dalmasso, F., Giordanengo, S., Monaco, V.,
  Muraro, S., Sacchi, R., Vignati, A., and Attili, A. (2017).
\newblock A monte carlo approach to the microdosimetric kinetic model to
  account for dose rate time structure effects in ion beam therapy with
  application in treatment planning simulations.
\newblock {\em Medical physics}, 44(4):1577--1589.

\bibitem[McMahon, 2018]{mcmahon2018linear}
McMahon, S.~J. (2018).
\newblock The linear quadratic model: usage, interpretation and challenges.
\newblock {\em Physics in Medicine \& Biology}, 64(1):01TR01.

\bibitem[Mein et~al., 2020]{mein2020assessment}
Mein, S., Klein, C., Kopp, B., Magro, G., Harrabi, S., Karger, C.~P., Haberer,
  T., Debus, J., Abdollahi, A., Dokic, I., et~al. (2020).
\newblock Assessment of rbe-weighted dose models for carbon ion therapy toward
  modernization of clinical practice at hit: in vitro, in vivo, and in
  patients.
\newblock {\em International Journal of Radiation Oncology* Biology* Physics},
  108(3):779--791.

\bibitem[Missiaggia et~al., 2020]{missiaggia2020microdosimetric}
Missiaggia, M., Cartechini, G., Scifoni, E., Rovituso, M., Tommasino, F.,
  Verroi, E., Durante, M., and La~Tessa, C. (2020).
\newblock Microdosimetric measurements as a tool to assess potential in-field
  and out-of-field toxicity regions in proton therapy.
\newblock {\em Physics in Medicine \& Biology}, 65(24):245024.

\bibitem[Missiaggia et~al., 2023]{missiaggia2023investigation}
Missiaggia, M., Cartechini, G., Tommasino, F., Scifoni, E., and La~Tessa, C.
  (2023).
\newblock Investigation of in-field and out-of-field radiation quality with
  microdosimetry and its impact on relative biological effectiveness in proton
  therapy.
\newblock {\em International Journal of Radiation Oncology* Biology* Physics},
  115(5):1269--1282.

\bibitem[Missiaggia et~al., 2022]{missiaggia2022cell}
Missiaggia, M., Cordoni, F.~G., Scifoni, E., and La~Tessa, C. (2022).
\newblock Cell survival computation via the generalized stochastic
  microdosimetric model ({GSM}2); part ii: Numerical results.
\newblock {\em (submitted)}.

\bibitem[Missiaggia et~al., 2021]{missiaggia2021novel}
Missiaggia, M., Pierobon, E., Castelluzzo, M., Perinelli, A., Cordoni, F.,
  Centis~Vignali, M., Borghi, G., Bellinzona, E., Scifoni, E., Tommasino, F.,
  et~al. (2021).
\newblock A novel hybrid microdosimeter for radiation field characterization
  based on the tissue equivalent proportional counter detector and low gain
  avalanche detectors tracker: a feasibility study.
\newblock {\em Frontiers in Physics}, 8:578444.

\bibitem[Monini et~al., 2019]{monini2019comparison}
Monini, C., Alphonse, G., Rodriguez-Lafrasse, C., Testa, {\'E}., and Beuve, M.
  (2019).
\newblock Comparison of biophysical models with experimental data for three
  cell lines in response to irradiation with monoenergetic ions.
\newblock {\em Physics and Imaging in Radiation Oncology}, 12:17--21.

\bibitem[Paganetti, 2014]{paganetti2014relative}
Paganetti, H. (2014).
\newblock Relative biological effectiveness (rbe) values for proton beam
  therapy. variations as a function of biological endpoint, dose, and linear
  energy transfer.
\newblock {\em Physics in Medicine \& Biology}, 59(22):R419.

\bibitem[Paganetti and Goitein, 2000]{paganetti2000radiobiological}
Paganetti, H. and Goitein, M. (2000).
\newblock Radiobiological significance of beamline dependent proton energy
  distributions in a spread-out bragg peak.
\newblock {\em Medical physics}, 27(5):1119--1126.

\bibitem[Perl et~al., 2012]{perl2012topas}
Perl, J., Shin, J., Sch{\"u}mann, J., Faddegon, B., and Paganetti, H. (2012).
\newblock Topas: an innovative proton monte carlo platform for research and
  clinical applications.
\newblock {\em Medical physics}, 39(11):6818--6837.

\bibitem[Rosenfeld, 2016]{rosenfeld2016novel}
Rosenfeld, A.~B. (2016).
\newblock Novel detectors for silicon based microdosimetry, their concepts and
  applications.
\newblock {\em Nuclear Instruments and Methods in Physics Research Section A:
  Accelerators, Spectrometers, Detectors and Associated Equipment},
  809:156--170.

\bibitem[Sato and Furusawa, 2012]{sato2012cell}
Sato, T. and Furusawa, Y. (2012).
\newblock Cell survival fraction estimation based on the probability densities
  of domain and cell nucleus specific energies using improved microdosimetric
  kinetic models.
\newblock {\em Radiation research}, 178(4):341--356.

\bibitem[Sato et~al., 2009]{sato2009biological}
Sato, T., Kase, Y., Watanabe, R., Niita, K., and Sihver, L. (2009).
\newblock Biological dose estimation for charged-particle therapy using an
  improved phits code coupled with a microdosimetric kinetic model.
\newblock {\em Radiation Research}, 171(1):107--117.

\bibitem[Sato et~al., 2013]{sato2013particle}
Sato, T., Niita, K., Matsuda, N., Hashimoto, S., Iwamoto, Y., Noda, S., Ogawa,
  T., Iwase, H., Nakashima, H., Fukahori, T., et~al. (2013).
\newblock Particle and heavy ion transport code system, phits, version 2.52.
\newblock {\em Journal of Nuclear Science and Technology}, 50(9):913--923.

\bibitem[Sato et~al., 2006]{sato2006development}
Sato, T., Watanabe, R., and Niita, K. (2006).
\newblock Development of a calculation method for estimating specific energy
  distribution in complex radiation fields.
\newblock {\em Radiation protection dosimetry}, 122(1-4):41--45.

\bibitem[Scholz et~al., 2001]{scholz2001direct}
Scholz, M., Jakob, B., and Taucher-Scholz, G. (2001).
\newblock Direct evidence for the spatial correlation between individual
  particle traversals and localized cdkn1a (p21) response induced by high-let
  radiation.
\newblock {\em Radiation research}, 156(5):558--563.

\bibitem[Tambas et~al., 2022]{tambas2022current}
Tambas, M., van~der Laan, H.~P., Steenbakkers, R.~J., Doyen, J., Timmermann,
  B., Orlandi, E., Hoyer, M., Haustermans, K., Georg, P., Burnet, N.~G., et~al.
  (2022).
\newblock Current practice in proton therapy delivery in adult cancer patients
  across europe.
\newblock {\em Radiotherapy and oncology}, 167:7--13.

\bibitem[Tommasino~et al., 2019]{tommasino2019new}
Tommasino~et al., F. (2019).
\newblock A new facility for proton radiobiology at the trento proton therapy
  centre: Design and implementation.
\newblock {\em Physica Medica}, 58:99--106.

\bibitem[Tran et~al., 2018]{tran2018relative}
Tran, L.~T., Bolst, D., Guatelli, S., Pogossov, A., Petasecca, M., Lerch,
  M.~L., Chartier, L., Prokopovich, D.~A., Reinhard, M.~I., Povoli, M., et~al.
  (2018).
\newblock The relative biological effectiveness for carbon, nitrogen, and
  oxygen ion beams using passive and scanning techniques evaluated with fully
  3d silicon microdosimeters.
\newblock {\em Medical physics}, 45(5):2299--2308.

\bibitem[Tran et~al., 2017]{tran2017characterization}
Tran, L.~T., Chartier, L., Bolst, D., Pogossov, A., Guatelli, S., Petasecca,
  M., Lerch, M.~L., Prokopovich, D.~A., Reinhard, M.~I., Clasie, B., et~al.
  (2017).
\newblock Characterization of proton pencil beam scanning and passive beam
  using a high spatial resolution solid-state microdosimeter.
\newblock {\em Medical physics}, 44(11):6085--6095.

\bibitem[Zaider et~al., 1996]{zaider1996microdosimetry}
Zaider, M., Rossi, B. H.~H., and Zaider, M. (1996).
\newblock {\em Microdosimetry and its Applications}.
\newblock Springer.

\bibitem[Zaider and Rossi, 1980]{zaider1980synergistic}
Zaider, M. and Rossi, H. (1980).
\newblock The synergistic effects of different radiations.
\newblock {\em Radiation research}, pages 732--739.

\bibitem[Zhu et~al., 2019]{zhu2019microdosimetric}
Zhu, H., Chen, Y., Sung, W., McNamara, A.~L., Tran, L.~T., Burigo, L.~N.,
  Rosenfeld, A.~B., Li, J., Faddegon, B., Schuemann, J., et~al. (2019).
\newblock The microdosimetric extension in topas: development and comparison
  with published data.
\newblock {\em Physics in Medicine \& Biology}, 64(14):145004.

\end{thebibliography}

\end{document}